\documentclass[acmsmall]{acmart}

\usepackage{xcolor}
\usepackage{color,soul}
\usepackage{amsmath}
\AtBeginDocument{%
  \providecommand\BibTeX{{%
    \normalfont B\kern-0.5em{\scshape i\kern-0.25em b}\kern-0.8em\TeX}}}

\definecolor{changecolor}{RGB}{176,48,96}

\setcopyright{acmlicensed}
\acmJournal{PACMHCI}
\acmYear{2024} \acmVolume{8} \acmNumber{CSCW1} \acmArticle{62} \acmMonth{4} \acmPrice{15.00}\acmDOI{10.1145/3637339}

\usepackage{hyperref}
\begin{document}

\title{Healthcare Voice AI Assistants: Factors Influencing Trust and Intention to Use}




\author{Xiao Zhan}
\orcid{0000-0003-1755-0976}
\affiliation{%
 \institution{King's College London}
 \country{United Kingdom}}
\email{xiao.zhan@kcl.ac.uk}

\author{Noura Abdi}
\orcid{0000-0002-4613-6443}
\affiliation{%
 \institution{Liverpool Hope University}
 \country{United Kingdom}}
  \email{abdin@hope.ac.uk}

\author{William Seymour}
\orcid{0000-0002-0256-6740}
\affiliation{%
\institution{King's College London}
\country{United Kingdom}}
\email{william.1.seymour@kcl.ac.uk}

\author{Jose Such}
\orcid{0000-0002-6041-178X}
\affiliation{%
 \institution{King's College London}
 \country{United Kingdom}
 \institution{\& VRAIN, Universitat Politecnica de Valencia}
 \country{Spain} 
 }
\email{jose.such@kcl.ac.uk}

\renewcommand{\shortauthors}{Zhan et al.}

\begin{abstract}

AI assistants such as Alexa, Google Assistant, and Siri, are making their way into the healthcare sector, offering a convenient way for users to access different healthcare services. Trust is a vital factor in the uptake of healthcare services, but the factors affecting trust in voice assistants used for healthcare are under-explored and this specialist domain introduces additional requirements. This study explores the effects of different functional, personal, and risk factors on trust in and adoption of healthcare voice AI assistants (HVAs), generating a partial least squares structural model from a survey of 300 voice assistant users. Our results indicate that trust in HVAs can be significantly explained by functional factors (usefulness, content credibility, quality of service relative to a healthcare professional), together with security, and privacy risks and personal stance in technology. We also discuss differences in terms of trust between HVAs and general-purpose voice assistants as well as implications that are unique to HVAs.
\end{abstract}


\setcopyright{acmlicensed}
\acmJournal{PACMHCI}
\acmYear{2024} \acmVolume{8} \acmNumber{CSCW1} \acmArticle{62} \acmMonth{4} \acmPrice{15.00}\acmDOI{10.1145/3637339}

\maketitle

\section{Introduction}
\label{introduction}


AI assistants such as Alexa, Google Assistant and Siri are now found in millions of smart speakers, phones, and other devices. Popular because of the hands-free, convenient experience they offer, voice assistants are mainly used to play music, search for information, and control smart IoT devices in the home~\cite{ammari2019music,sciuto2018hey,dambanemuya2021auditing}. However, the capabilities of voice assistants are continuously expanding to cover other uses and enahnce user experience. Capabilities added by third parties (often called `skills' or `actions') continue to grow, with the number of Alexa skills surpassing 100,000~\cite{alexa-skills}. Beyond the main uses mentioned above, there are many other popular skill categories including ride hailing (e.g. Uber), checking and sending emails (e.g. Myemail), and online banking.



One particular domain that is gaining attention is the application of voice assistants in healthcare, which we refer to as Healthcare Voice AI Assistants (HVAs)~\cite{voicebot2022,pradhan2020use}. In particular, HVAs offer more and more healthcare services such as virtual assessments for patients, diagnostic suggestions, and health and lifestyle tips \cite{alexahealth}. 
One of the key advantages of HVAs is that they are a cost-effective alternative to in-person or telephone consultations, especially at a time when the number of healthcare professionals is not enough to meet the needs of patients~\cite{9551864}. They also offer convenience and efficiency when patients need to travel for long distances to see a doctor~\cite{9551864}, or when situations like the Covid-19 pandemic make it difficult for patients to be attended to physically~\cite{bharti2020medbot}. In fact, many healthcare providers around the world are already using HVAs in some way, such as the UK's National Health Service which offers information via Alexa.\footnote{\url{https://www.gov.uk/government/news/nhs-health-information-available-through-amazon-s-alexa}} The number of US citizens that used voice assistants for healthcare related purposes has nearly tripled since 2019.\footnote{See the report \url{https://voicebot.ai/2022/01/07/voice-assistant-use-in-healthcare-nearly-tripled-over-two-years-and-demand-still-outstrips-supply-new-report/} for details.} Beyond what is currently available through contemporary voice assistants, researchers have been exploring ways to deliver a wide range of healthcare services remotely through HVAs~\cite{wolters2016designing, langevin2019park}. However, HVAs also pose challenges. Privacy and data security are critical due to the sensitivity of health information involved~\cite{price2019privacy} and the absence of robust AI healthcare regulations potentially magnifies these issues, inviting misuse or data misinterpretation~\cite{o2019legal}. There is also a risk of inaccuracies in HVA diagnoses and advice, potentially leading to poor health outcomes~\cite{abbasi2006clinical}. It is therefore crucial to address these issues as the field progresses.





As HVAs continue to develop and improve, a question arises as to whether people will trust and adopt them. The multifaceted nature of trust in both technology and general-purpose voice assistants has been extensively studied 
(see Section~\ref{related-work} for a comprehensive review). This body of work has substantiated the correlation between trust and user acceptance~\cite{alhogail2018improving,lu2011dynamics,sharma2020consumer,liao2019understanding,pitardi2021alexa,al2022understanding,nasirian2017ai,lee2021role}, and includes comprehensive studies across diverse domains on the effect of functional~\cite{mcknight2011trust,waytz2014mind,chang2017user,peng2020patient,xu2014different,mcknight2011trust,cho2009effect,chang2013service}, personal~\cite{mcknight2011trust,chi2021developing,chi2021developing,alhogail2018improving}, and risk~\cite{mayer1995integrative,pavlou2003consumer,ha2020effects,egea2011explaining} factors in trust, thereby reinforcing the paramount importance of trust within the technological sphere. However, there is a need to study trust specifically in the context of HVAs because the healthcare domain is known to bring a range of additional challenges~\cite{bickmore2018patient,asan2020artificial}. For instance, despite the shared influence of certain factors, such as the \emph{reliability of services and safety}, it is evident that human trust in healthcare and trust in other technologies exhibit notable disparities in their respective levels of confidence~\cite{montague2009empirically}. As described above, healthcare data is also more sensitive than other types of data, as shown by previous academic works~\cite{miner2016smartphone,abdi2021privacy}, and is often subject to additional protection under privacy regulations such as the GDPR in Europe\footnote{The General Data Protection Regulation (GDPR) in the EU designates healthcare data as `special category', requiring additional protections (Article 9 GDPR) \url{https://gdpr.eu/tag/gdpr/}} and HIPPA in the United States.\footnote{The Health Insurance Portability and Accountability Act of 1996 (HIPAA) sets the standard for sensitive patient data protection \url{https://www.hhs.gov/hipaa/index.html}.}
%
Furthermore, it has been shown that trust in traditional healthcare providers such as physicians~\cite{hall2001trust,ward2018trust, gregory2021patients} cannot be directly applied to services delivered by HVAs~\cite{wang2018living}. This is because interacting with a voice assistant is fundamentally distinct from interacting with a human, even if voice assistants exhibit several human-like characteristics (like voice interaction)~\cite{aeschlimann2020communicative}. In addition, AI algorithms embedded in HVAs lead to decision-making processes that may not be apparent to the users or that are not adequately explained compared to healthcare professionals
~\cite{gille2020we,vollmer2018machine,nagendran2020artificial}. Finally, issues related to aspects that affect trust may be more pre-eminent. For instance, incomplete or incorrect health-related information, such as first-aid instructions or medication recommendations may cause patients harm~\cite{sezgin2020readiness}. 




As a field, CSCW has shown a keen interest in digital healthcare in particular studying its socio-technical dimensions. This encompasses the collaborative efforts inherent in healthcare AI and the considerations steered by policy perspectives~\cite{fitzpatrick2013review,park2019identifying}. 
The study of patient trust in Healthcare AI in particular aligns with these core themes. Besides, 
Healthcare AI is a multidisciplinary field, from medical professionals, AI developers, data scientists and other key stakeholders 
~\cite{lai2021human,park2019identifying}. Their collective efforts are needed for the development, implementation, and refinement of healthcare solutions, in particular those focusing on the role of trust. There are also CSCW studies examining communication dynamics between patients and providers~\cite{park2017beyond}. These studies emphasize the importance of integrating diverse stakeholder perspectives to foster a holistic understanding of the relationship between patients and providers. Equally crucial is 
the unique situations and expectations of specific stakeholder groups
, like 
the shared experiences and challenges faced by patients, 
forming a cohesive community actively engaging with AI technologies. 

Given that the healthcare sector serves as a vital domain for human-AI collaboration~\cite{lai2021human} and considering trust as the cornerstone fostering this collaboration~\cite{drozdal2020trust,arnold2019factsheets,wang2020human,jirotka2005collaboration,10.1145/3500868.3559450}, it is vital to extensively research and investigate in order to further enhance and supplement this specific research field. The socio-technical implications of HVAs, which bridge the gap between technology and social healthcare practices~\cite{fitzpatrick2013review}, further emphasize the importance of understanding trust dynamics in this context. Finally, the two central tenets delineated in our paper, `trust' and `adoption', inherently embody a collaborative essence~\cite{lampinen2016cscw,muller2016collaborative}. This dual-faceted approach not only seeks to enhance user sentiment and satisfaction but also addresses the developmental requirements discernible to HVA developers.
To the best of our knowledge, this work is the first to examine factors that affect trust and intention to use in HVAs. We study the effect of different functional (Anthropomorphism, Effort Expectancy, Perceived Usefulness, Perceived Content Credibility, and Perceived Relative Service Quality), personal (Stance in Technology, Familiarity, Technology Attachment, and Social Influence), and risk (Security Risk, Privacy Risk, and Perceived Substitution Risk) factors on trust in HVAs and, in turn, the effect of this trust on peoples' intention to use HVAs. The selection of our factors is based on those with a confirmed substantial impact on trust in broader contexts such as general voice assistants, technology at large, and the healthcare industry, in order to measure their effect in the context of HVAs (as we detail in Section~\ref{sec: Hypothesis}). More specifically, this study aims to answer the following research questions: 

\begin{description}
    
    \item[RQ1] How do functional, personal and risk factors influence users' trust in healthcare voice assistants?
    \item[RQ2] How does trust in healthcare voice assistants influence users' intention to use them?
\end{description}

To answer these questions, we developed a model of factors affecting trust in HVA and intention to use. In order to test the model we administered a questionnaire to 300 participants which included questions measuring these latent functional, personal, and risk variables alongside trust and intention to use. The results were analysed using partial least squares structural modelling (PLS-SEM) and show how functional factors (anthropomorphism, perceived usefulness, effort expectancy, content credibility, and relative service quality), risk factors (security risk and privacy risk), and personal factors (stance in technology) substantially explain trust in HVAs. They also suggest that trust in HVAs explains intention to use HVAs. Based on these results we derive some practical insights and recommendations to help design and develop the next-generation of HVAs, promoting user trust and encouraging adoption. 

\section{Related Work}
\label{related-work}


\subsection{Trust in Technology}
\label{trust_frameworks}

\subsubsection{Trust Scales and Measurement}
The topic of trust has been extensively investigated, but there is not yet a completely agreed definition of what it is and what it means. 
Trust has often been described as "a willingness to ascribe good intentions to and have confidence in the words and actions of other people" in social science~\cite{fox1974beyond}, while trust in technology usually refers to "beliefs that a specific technology has the attributes necessary to perform as expected in a given
situation in which negative
consequences are possible."~\cite{mayer1995integrative}. Trust is recognized as a central focus for the CSCW discipline 
over the past several decades~\cite{lampinen2016cscw}. Trust is recognized as a pivotal factor influencing human-AI collaboration~\cite{jirotka2005collaboration,10.1145/3500868.3559450}, and in particular, previous work published at CSCW~\cite{jirotka2005collaboration} has underscored the positive effects of research on trust in fostering human-AI collaboration. 
In order to study the different ways in which other factors can affect trust, a critical step is to be able to measure trust in a particular technology, in our case, HVA. Previous literature proposed different metrics and scales to measure trust in technology~\cite{muir1996trust,mayer1995integrative,benbasat2005trust,koh2010effects}. More recently, and in the context of general-purpose voice assistants such as Amazon Alexa, \citeauthor{cho2020will}~\cite{cho2020will} used a scale to measure trust, adapting it from \cite{koh2010effects}, where they consider users' perceptions about the assistant's competence, integrity, and benevolence, which were suggested by previous studies to be key components of trust ~\cite{muir1996trust,mayer1995integrative,benbasat2005trust,koh2010effects}. As we explain later on, we adapt the scale proposed in \cite{cho2020will} to measure trust in our study.  

\subsubsection{Adoption} 

As studied in early CSCW literature~\cite{muller2016collaborative}, the dynamics of collaborative work are often embedded in the processes of technology adoption and adaptation. This suggests that even if the focus seems individualistic, the broader collaborative context is often at play. In terms of the models to study adoption, 
one of the most well-known and used models is the Technology Acceptance Model (TAM). The model was first proposed by Davis in 1989, and it suggests that an individual's acceptance of technology is determined by two key factors: perceived usefulness and perceived ease of use. Subsequent models built on TAM (e.g. TAM2~\cite{venkatesh2000theoretical}, TAM3~\cite{venkatesh2008technology}, and UTAUT~\cite{venkatesh2003user}) have extended the original TAM by incorporating additional variables such as social influence~\cite{venkatesh2000theoretical,venkatesh2008technology,venkatesh2003user}, system characteristics~\cite{venkatesh2008technology}, and facilitating conditions~\cite{venkatesh2008technology,venkatesh2003user} to explain the acceptance and use of technology. However, recent works~\cite{lowe2019consumers,mclean2019hey} have criticised their effectiveness as being highly context-dependent, and that they are too simplified for emerging technologies. It is therefore now common to combine them with other theories and attributes that influence users' usage behaviour, as we elaborate in the next section,  
to compensate for their limitations in focusing more on some of the functional factors.

\subsubsection{Trust and Adoption} \label{sec: trust and VA}


Investigating users’ trust in new technologies and products is of great significance, because user 
acceptance and adoption are substantially and significantly influenced by 
trust~\cite{alhogail2018improving,lu2011dynamics,sharma2020consumer,liao2019understanding,pitardi2021alexa,al2022understanding,nasirian2017ai,lee2021role}. In particular, in all of these studies, trust has shown a positive effect on users' behavioural intentions to use a technology in domains such as e-commerce~\cite{sharma2020consumer}, mobile payments~\cite{lu2011dynamics}, and IoT devices~\cite{alhogail2018improving}. This has also been shown recently for general-purpose voice assistants~\cite{nasirian2017ai,lee2021role,liao2019understanding,pitardi2021alexa,al2022understanding}.
%
%
For instance, 
Liao et al.~\cite{liao2019understanding} found that users' attitudes towards adoption were rooted in whether they trusted voice assistants
, and 
Pitardi et al.~\cite{pitardi2021alexa} showed that trust is one of the key contributors to foster adoption of 
Google Assistant. 
More recently, the importance of trust for adoption of voice assistants was also shown when assistants are used for specific tasks, such as in an education setting
~\cite{al2022understanding}.


\subsection{Factors Affecting Trust} \label{sec:factor-literature}


HCI studies examining how different factors affect users' trust in technology have identified three main types of factors that affect trust, including functional, personal and risk factors. Functional factors refer to the characteristics of the technologies themselves~\cite{mcknight2011trust,waytz2014mind,chang2017user,peng2020patient,xu2014different,mcknight2011trust,cho2009effect,chang2013service}, personal factors refer to the individual user, their social circle and their experiences~\cite{mcknight2011trust,chi2021developing,chi2021developing,alhogail2018improving}, and risk factors refer to the harms that could be derived from the technology~\cite{mayer1995integrative,pavlou2003consumer,ha2020effects,egea2011explaining}. In a related vein, Knowles et al. \cite{knowles2015models} emphasized trust as crucial for collaboration and introduced design patterns to enhance trust. While not empirically validated like the aforementioned models, these patterns highlight the importance of factors such as usability (functional factor) and socialization (personal factor).

Regarding the functional factors that affect trust in technology, previous literature has uncovered factors such as the usefulness of the technology and how easy the technology is to use~\cite{chandra2018analysis}, as well as its reliability~\cite{lewandowsky2000dynamics,madhavan2006automation} and the interfaces designed for interaction ~\cite{parasuraman2004trust}. These all manifest slightly differently in voice assistants, particularly interface design and reliability. With regards to interface design,  Anthropomorphism~\cite{chen2021anthropomorphism,seymour2021exploring,van2019trust} has been shown to influence trust, with studies suggesting that voice alone facilitates 
human connection with the assistant~\cite{han2018understanding,novak2019relationship}. Regarding reliability, and partly because of how voice assistants operate, the broader concept of content credibility~\cite{lou2019influencer}, which includes reliability as well as other aspects such as accuracy, completeness and authenticity of the information, is a critical factor to foster trust in voice assistants, that is, the credibility of what the voice assistant utters back to the user seems to affect how much the user trusts the assistant. 

In addition to functional elements, whether users consider technology to be trustworthy depends on personal factors. This includes individual factors and social factors. Individual factors, such as user predisposition to try new technologies~\cite{agarwal1998conceptual}, user familiarity with an understanding of a technology~\cite{komiak2006effects,chi2021developing}, and even the 
users' emotional attachment~\cite{you2017emotional} and psychological connection to a technology~\cite{suh2011if}. 
Social factors also seem to play a role, with works like~\cite{gursoy2017preferences} showing that social influence, the degree that a customer's social group
(e.g. family, friends, etc.) believes that using a technology is relevant, affects trust in that technology. 




Finally, the effect of risk on trust has long been studied~\cite{mayer1995integrative}, and shown to manifest across different domains, including 
online transactions~\cite{pavlou2003consumer}, autonomous vehicles~\cite{ha2020effects}, and the use of Electronic Health Care Records (EHCR) by physicians~\cite{egea2011explaining}. 
Security and privacy risks, in particular, have been shown to reduce the benefits of using general-purpose voice assistants~\cite{mclean2019hey,kowalczuk2018consumer,alhogail2018improving}, directly impacting the trust users' have in voice assistants, with users stopping using voice assistants all together as a result, or restricting their use of voice assistants only to functionality they perceive to be less risky~\cite{bansal2016context,abdi2019more}. 





\subsection{Healthcare Voice Assistants}


\subsubsection{HVA available in Commercial Voice Assistants}
Commercial voice assistants nowadays incorporate healthcare capabilities as Amazon Alexa \emph{Skills} or Google Home \emph{Actions}~\cite{chung2018health}. 
In 2020, researchers grouped the spectrum of healthcare services provided by voice assistants into four levels: information-, assistance-, assessment-, and support-level services~\cite{sezgin2020scoping,sezgin2020capturing}. Of these, information-level services (providing healthcare-related information to users) are one of the most common and basic functions and are already widely embedded in most commercially available voice assistants. There are many Alexa skills, for example, that provide healthcare information such as descriptions/or instructions on symptoms, medications, and how to use medical facilities (e.g., Dr. A.I., WebMD, and Mayo Clinic). 
Examples of assistance-level services include the ability for VAs to assist users in setting up reminders for medication or physical activity (i.e. tracking their symptoms) and providing appointment notifications~\cite{sezgin2020scoping,sezgin2020capturing}. Some of these features are also available in contemporary VAs like the My Children's Enhanced Recovery After Surgery skill developed by Boston Children's Hospital, which reminds parents and caregivers about information regarding their post-op appointments. 
%
%
%
%
%
Support for functionality within the assessment and support layers ranges from scarce to non-existent. Assessment level skills, for example, would enable voice assistants to recognise changes in mood or health conditions, but commercial voice assistants are only currently able to access and view diagnostic results and records (e.g. Alexa Skill Sugarmate). Similarly, the support level includes the ability to prescribe and prioritise care/treatment but users have thus far only been able to refill their prescriptions\footnote{This feature was available for Alexa owners in 2019 - \url{https://www.dailymail.co.uk/sciencetech/article-7727961/Alexa-owners-use-virtual-assistant-refill-prescriptions.html}}. 




\subsubsection{Research on HVAs}
Extensive research has been conducted recently on the use of voice assistants for healthcare. 
For example, \citeauthor{dojchinovski2019interactive} \cite{dojchinovski2019interactive} saw the potential for VAs to become an important part of the healthcare system for heart disease. The VAs could communicate directly with the user through their special voice interface and then, retrieve the required information from the medical cloud and transmit it back to the user. In this case, users would highly benefit from using a VA-based healthcare system to perform tasks such as checking ECG readings, scheduling appointments, recording treatments and contacting doctors. Other strategies and prototypes for coping with various healthcare issues have been discussed as well, such as supporting home care~\cite{10.1145/3491102.3517683}, and improving patients' Type2 diabetes management~\cite{buinhas2018virtual,cheng2018development}. 
Another stream of research on HVA has focused on evaluating the performance of voice assistants in healthcare-related tasks. Google Assistant, Amazon Alexa, Apple Siri, Microsoft Cortana and other voice commercial assistants are often tested and compared ~\cite{alagha2019evaluating, kocaballi2020responses, nobles2020responses, yang2021clinical}. Overall, the ability of VAs to provide health-related advice is very limited and performed unfavourably to identify the user's queries for healthcare information
~\cite{alagha2019evaluating,nobles2020responses,yang2021clinical,kocaballi2020responses}. 
Finally, there has been some recent work examining how
differences in modality (i.e., voice vs. text) and device type (i.e.,
smartphone vs. smart home device) affect user perceptions
when retrieving sensitive health information from voice assistants~\cite{cho2019hey}, finding that voice led to more social presence and that device did not matter. In addition, attention was paid to the behaviour and perception of older people using voice assistants for the specific purpose of seeking healthcare information~\cite{zhong2022effects,brewer2022empirical}. They found that older people's evaluation of the VA was influenced by communication styles, anthropomorphic settings, and individual difference~\cite{zhong2022effects}, and that they appeared to use voice assistants primarily to confirm information obtained from other sources (e.g., laptop, phone)~\cite{brewer2022empirical}. 
While previous work~\cite{park2019identifying} 
emphasized the importance of addressing trust issues, especially amidst evolving healthcare workflows, we are unaware of any prior research specifically targeting user trust in HVAs and the influencing factors. 

\section{Hypotheses}
\label{sec: Hypothesis}

\begin{table}[tbh]
    \centering
    \small
    \caption{Summary of factors considered in the model. (H1-5 are functional factors, H6-9  personal factors, and H10-12  risk factors.)}
    \scalebox{0.91}{
    \begin{tabular}{p{0.15cm}p{2.7cm}p{8.3cm}p{2.5cm}}
    \toprule
    H\# &Factors & Description & Supporting Source \\
    \midrule
    H1& Anthropomorphism &Anthropomorphism in voice assistants involves human-like characteristics, creating a social connection with users. Voice-based interaction deepens this bond between users and the assistant. & \cite{moussawi2021perceptions,epley2007seeing,han2018understanding,novak2019relationship,van2017domo,pitardi2021alexa,tanioka2021development}\\
    H2& Effort Expectancy &Effort expectancy is the predicted mental and physical activity required to use a technology. It affects trust in technology, and challenges in ease of use can lead to anxiety and distrust. &\cite{venkatesh2012consumer,pitardi2021alexa,alhogail2018improving,kulviwat2007toward,davis1989perceived,lin2020antecedents,chang2017user,vaportzis2017older,lee2015perspective,10.1145/3196490,chi2021developing,schnall2015trust} \\
    H3& Per. Usefulness &is about device efficiency in daily tasks. In healthcare, it means technology facilitating services. Usefulness affects attitudes and adoption. In HVAs, useful responses to symptoms are vital for user satisfaction. &\cite{buteau2021hey,davis1989perceived,coughlan2012exploring,gao2014unified,alhogail2018improving,peng2020patient,lee2021application} \\
    H4& Per. Content Credibility & Content credibility is crucial for user trust. In healthcare, users rely on accurate information from virtual assistants. Inaccurate information can be harmful.& \cite{xu2014different,mcknight2011trust,cho2009effect,lewandowsky2000dynamics,madhavan2006automation,asan2020artificial,sunarsi2019effect,cho2020will,appelman2016measuring,bickmore2018patient,nadarzynski2019acceptability,haan2019qualitative,sezgin2020readiness}.\\
    H5& Per. Relative Service Quality & This perception compares healthcare technology to professionals. Patients value professional care but may trust technology for specific diagnoses if it offers greater accuracy.& \cite{baker2020comparison,young2021patient,liu2019comparison,promberger2006patients,haan2019qualitative,jutzi2020artificial}\\
    \midrule
    H6& Familiarity &Familiarity represents the level of understanding of an object. When users are familiar with a technology, they are more likely to trust it compared to unfamiliar ones in various domains such as e-commerce, IoT devices, and voice assistants. & \cite{komiak2006effects,luhmann2001familiarity,gefen2004consumer,alraja2019effect,chi2021developing}\\
    H7& Tech Attachment & This factor, aka emotional attachment, reflects the psychological connection between individuals and a technology.& \cite{you2017emotional,suh2011if,perlaviciute2014contextual,chi2021developing}\\
    H8& Social Influence &Social influence plays a significant role in shaping users' trust and perceptions of a product, as they are influenced by the behaviors and opinions of others in their social networks. &\cite{latane1981psychology,gursoy2017preferences,tajfel1979integrative,alhogail2018improving} \\
    H9& Stance in Tech & Stance in technology reflects an individual's openness to trying new innovations, with a positive correlation found between willingness to experiment with new technologies and acceptance and trust in their services.& \cite{agarwal1998conceptual,mcknight2011trust,chi2021developing}\\
    \midrule
    H10& Security Risk &Security risk in technology, especially in healthcare, can undermine trust due to potential consequences of flaws and incidents such as errors, malware, and hacking. &\cite{alraja2019effect,keskinbora2019medical,ellahham2020application,lee2021included,nadarzynski2019acceptability,haan2019qualitative} \\
    H11& Privacy Risk &Privacy risk arises when users perceive a potential threat of data misuse or unauthorized access, particularly in the sensitive healthcare domain, impacting trust in voice assistants. &\cite{alraja2019effect,pitardi2021alexa,vimalkumar2021okay,foehr2020alexa,hayden2013privacy,topol2019high,gymrek2013identifying,easwara2015privacy,miner2016smartphone,abdi2021privacy} \\
    H12& Per. Substitution Risk & Perceived substitution risk in healthcare technologies involves the concern of technology replacing traditional healthcare professionals, impacting trust in AI applications.& \cite{fan2020investigating,ongena2020patients,haan2019qualitative,nadarzynski2019acceptability,jutzi2020artificial,nelson2020patient}\\
    \midrule
    H13& Intention to Use &Intention to use reflects the connection between trust and technology adoption, with a strong correlation between users' trust, attitudes, and actual use of technology. & \cite{alhogail2018improving,lu2011dynamics,sharma2020consumer,nasirian2017ai,lee2021role,pitardi2021alexa,liao2019understanding,al2022understanding,chao2019factors}\\
    \bottomrule
    \end{tabular}}
    \label{tab:my_label}
\end{table}

%
%

\begin{figure}[h]
    \centering
    \includegraphics[width=\textwidth]{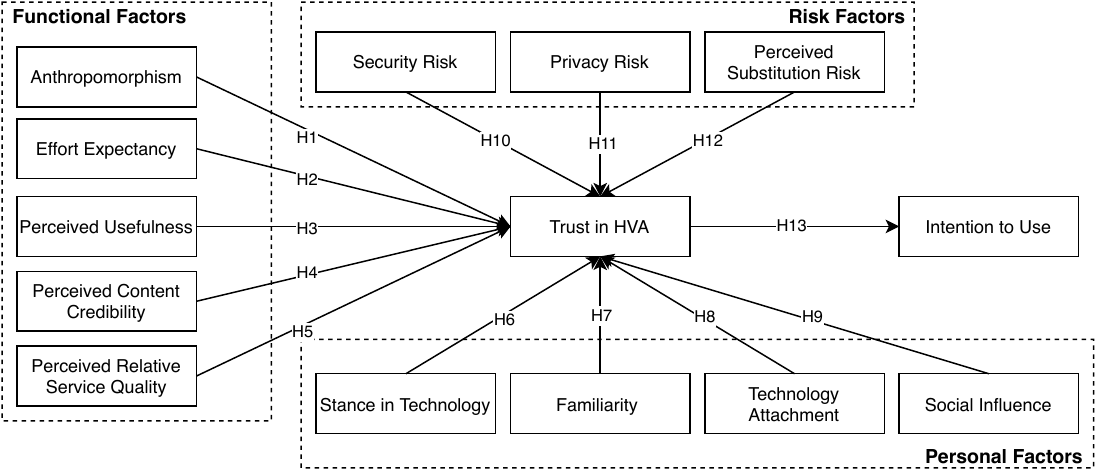}
    \vspace{-5pt}
    \caption{Hypothesized Research Model} 
    \label{fig:hypothesis model}
\end{figure}

Trust is a complex phenomenon that can be affected by many different factors. As discussed in Section~\ref{sec:factor-literature}, there are a number of personal, functional, and risk factors that have been shown to have an impact on users' trust in technology. Therefore, in this paper, we consider factors belonging to these categories that were shown by previous works to have a significant effect on trust adapted to the case of HVAs. In particular, and as we detail below, we consider: i) factors that have been shown to affect trust on general-purpose voice assistants such as Amazon Alexa and Google Assistant~\cite{pitardi2021alexa,chi2021developing,alhogail2018improving,foehr2020alexa,alraja2019effect,vimalkumar2021okay} ; ii) factors that have been shown to affect trust in technology in general~\cite{waytz2014mind,chang2017user,cho2009effect,xu2014different,gefen2004consumer}; and iii) factors that are specific to trust in the healthcare domain~\cite{peng2020patient,chang2013service,yokoi2021artificial}. We do not include factors that, while significant in prior works are not relevant or do not translate to the case of HVAs. For instance, \textit{Hedonic Benefits} is one of the factors that seem to affect trust in general-purpose voice assistants~\cite{mclean2019hey,pitardi2021alexa}, but that only applies when the purpose of use is hedonistic~\cite{wu2010falling}, and it is not aligned with the relatively more serious, purposeful, and specific context of seeking health care. Also, we are interested in factors that affect trust rather than the measurement of trust as a construct~\cite{alhogail2018improving}, so we adapt already existing scales to measure trust~\cite{cho2020will} and focus on the factors that influence it. 
Figure~\ref{fig:hypothesis model} summarizes our hypothesized research model, the different factors that may affect trust in HVA and its relationship to intention to use HVA. We now detail the categories, factors, and our hypotheses.

\subsection{Functional Factors}
The extensive literature on trust in technology  identified and subsequently confirmed across different domains that users' trust in a technology is influenced by whether the technology delivers on the functionality promised to complete the tasks or services they are supposed to offer~\cite{mcknight2011trust,yan2008trust}. This normally refers to both design and performance issues~\cite{yan2008trust}. When talking about voice assistants, functional factors take particular manifestations, and have been thoroughly examined both qualitatively and quantitatively in the  previous study~\cite{dambanemuya2021auditing,seymour2021exploring,cho2020role}. For instance, and as further detailed below, because of the very nature of their interaction modality, they offer a more anthropomorphic experience than other technologies. Another example, and in particular when talking about HVA, is that users may compare the healthcare service offered by an HVA relative to what users normally get from healthcare professionals. 




\subsubsection{Anthropomorphism (A)}

This factor refers to the extent that voice assistants show human-like characteristics and are therefore perceived and treated by users in ways that may be closer to how they would interact with another human~\cite{moussawi2021perceptions}. The more human-like characteristics voice assistants display the more it seems humans connect with them~\cite{epley2007seeing}. Starting with the modality of interaction, voice alone facilitates 
human connection with the assistant~\cite{han2018understanding,novak2019relationship}, and it may make users feel they are interacting with a social entity~\cite{van2017domo}. In fact, previous research on general-purpose voice assistants showed that their anthropomorphic characteristics have a positive effect on users' trust in voice assistants~\cite{van2017domo,pitardi2021alexa}. We hypothesize this to also be the case in HVA, particularly as healthcare is a domain where human-like characteristics in the technology used may contribute to a satisfactory experience~\cite{tanioka2021development}. 

\textbf{H1} Anthropomorphism positively influences Trust in HVAs.
    
\subsubsection{Effort Expectancy (EE)} This factor is the  predicted mental and physical activity and skill required to use a technology~\cite{venkatesh2012consumer}. In general, it has been shown across different domains that effort expectancy, or its flip side `ease of use', affect trust in technology~\cite{pitardi2021alexa,alhogail2018improving,kulviwat2007toward,davis1989perceived,lin2020antecedents,chang2017user}. For instance, one case is that of older adults, who find it difficult to use digital technologies~\cite{vaportzis2017older,lee2015perspective}, which has been shown to trigger anxiety and distrust~\cite{lee2015perspective,10.1145/3196490}. There is no previous work on the relationship of effort expectancy and voice assistants (not even in general-purpose ones). However, given the negative role that effort expectancy has been shown to play in similar technologies before, such as social robots for service delivery~\cite{chi2021developing}, we hypothesize that effort expectancy also plays a role in trust in HVA, particularly it can negatively affect it. This would also be consistent with work on mobile healthcare technologies that showed that the reverse of effort expectancy has a positive effect on intention to use~\cite{schnall2015trust}.

\textbf{H2} Effort expectancy negatively influences Trust in HVAs.

\subsubsection{Perceived Usefulness (PU)}

In general, this factor refers to "the degree to which the device makes life more efficient and helps carry out daily tasks."~\cite{buteau2021hey}. In other words, and particularly considering the healthcare domain, this factor refers to whether and how technology helps users to seek and get healthcare services in an efficient and effective way. 
According to \citeauthor{davis1989perceived}\cite{davis1989perceived}, usefulness can influence  attitudes towards a particular technology, and in particular user acceptance~\cite{coughlan2012exploring,gao2014unified} and adoption~\cite{alhogail2018improving} of technologies. The contribution of usefulness to trust has been demonstrated, especially in the healthcare domain. For instance, \citeauthor{peng2020patient}~\cite{peng2020patient} confirmed that the usefulness of online information and suggestions provided by physicians has a positive effect on patient-physician trust. Similarly, when using HVAs, the usefulness of responses from the HVA to user queries describing symptoms is suggested to be of great significance~\cite{lee2021application}, and it might cause differences in the user's feelings and judgements. Based on all of this, we formulate the following hypothesis: 



\textbf{H3} Perceived Usefulness positively influences Trust in HVAs.

\subsubsection{Perceived Content Credibility (PCC)}
This factor represents the credibility of the information provided by the technology, and it is often identified as a key factor influencing users' trust in 
technology in general~\cite{xu2014different,mcknight2011trust,cho2009effect}. The credibility of information is known to, in turn,   be composed of different aspects, such as its accuracy, reliability (including completeness, consistency, and authenticity), and authoritativeness (the information was generated by experts)~\cite{lewandowsky2000dynamics,madhavan2006automation,asan2020artificial,sunarsi2019effect,cho2020will,appelman2016measuring}. These are all critical aspects of information in the healthcare context because users may blindly follow the instructions and advice given by HVAs, such as what medication to take and how often to take it~\cite{bickmore2018patient}. Inaccurate information is therefore likely to result in the user being harmed~\cite{bickmore2018patient,nadarzynski2019acceptability,haan2019qualitative,sezgin2020readiness}. 

\textbf{H4} Perceived Content Credibility positively influences Trust in HVAs.

\subsubsection{Perceived Relative Service Quality (PRSQ)} 
This factor refers to the quality of a healthcare service a technology offers \emph{compared} to the same service offered by a healthcare professional. It has been used by previous research as an important indicator of the satisfaction healthcare technologies can bring compared to traditional healthcare services~\cite{baker2020comparison,young2021patient,liu2019comparison}. 
%
%
When it comes to trust, research showed that patients are reluctant to trust AI technologies because they value the quality of care provided by healthcare professionals~\cite{promberger2006patients}. In addition, qualitative studies on trust in AI technologies for particular types of diagnoses (like cancer)~\cite{haan2019qualitative,jutzi2020artificial} suggested that some patients might trust the technology if it provides more accuracy than healthcare professionals. We therefore hypothesize that PRSQ affects trust and that the higher the PRSQ (that is the higher the perception of quality of the service when compared with a healthcare professional) the higher the trust.

\textbf{H5} Perceived Relative Service Quality positively influences Trust in HVAs.


\subsection{Personal Factors}



Users may have differing attitudes toward the same technology~\cite{merritt2008not,lee2001trust,10.1145/3359161}. Trust is no exception, and it can be affected by individual and social factors. For instance, an individual may be more receptive to new technology than others (e.g. early adopters), and another individual may also (or instead) be influenced by the  opinions of their (online and/or offline) social circle. 
Therefore, in addition to functional factors, we also consider personal factors.

\subsubsection{Familiarity (F)} 
This factor refers to the degree of understanding of an object~\cite{komiak2006effects}. Research has shown that the more familiar users are with a technology the more likely they are to trust it in comparison to other technologies they do not know about~\cite{luhmann2001familiarity}. In fact, it has been shown to increase trust by reducing users' uncertainty on the technology in e-commerce~\cite{gefen2004consumer}, IoT devices~\cite{alraja2019effect}, and voice assistants~\cite{chi2021developing}. 
Similarly, in this study, we hypothesize that familiarity with HVA influences the trusts users have on them~\cite{chi2021developing}. 

\textbf{H6} Familiarity positively influences Trust in HVAs.

\subsubsection{Technology Attachment (TA)} This factor, also known as emotional attachment~\cite{you2017emotional}, represents the psychological connection between a person and a technology~\cite{suh2011if}. 
This connection is known to promote a positive view of the technology in question~\cite{perlaviciute2014contextual}. In particular, in general-purpose voice assistant, TA is known to significantly affect trust~\cite{chi2021developing}. 

\textbf{H7} Technology Attachment positively influences Trust in HVAs.


\subsubsection{Social Influence (SI)} Users are influenced by the behaviors and opinions of others in their social networks, according to the theory of social influence~\cite{latane1981psychology}. They tend to believe the same as the majority in those networks~\cite{gursoy2017preferences}. This is consistent with Social Identity Theory ~\cite{tajfel1979integrative}, which suggests that following social norms makes people feel integrated. As a result, users' perceptions, and in particular their trust towards a product, can vary~\cite{alhogail2018improving}, based on the trust their social connections have. 
In this study, we hypothesize that this factor reflects the propensity of users to trust HVAs, to capture the influence their social network may have on it. 

\textbf{H8} Social Influence (positively or negatively) influences Trust in HVAs.

\subsubsection{Stance in Technology (ST)} This factor is the extent to which an individual would be open to try out a new innovation or technology~\cite{agarwal1998conceptual}. Many existing studies have found that people who are more willing to experiment with and try new technologies are more inclined to accept and trust the services that these technologies offer~\cite{mcknight2011trust,chi2021developing}. It is therefore reasonable to hypothesize that such a group of individuals would likely be  open towards HVAs. 


\textbf{H9} Stance in technology positively influences Trust in HVAs.


\subsection{Risk Factors}
The use of technologies usually comes with some risks~\cite{lau2018alexa,malkin2022can,meng2021owning,mcmillan2019designing,seymour2023ignorance,10.1145/3359161}. These risks may also affect the trust that users have in the technologies. For instance, if the perceptions of risk is high, research has shown that that can lead to distrust~\cite{koien2011reflections,alhogail2018improving}. We consider the following risk factors and the effect they may have on trust in HVA.

\subsubsection{Security Risk (SR)} 
This factor relates to how unsafe and unprotected users feel when using a technology. Previous research on other healthcare technologies showed that SR is negatively associated with trust~\cite{alraja2019effect}. This is, in part, because healthcare is a critical domain, where flaws in the technology can lead to serious issues~\cite{keskinbora2019medical,ellahham2020application,lee2021included,nadarzynski2019acceptability,haan2019qualitative}. In addition, the digital healthcare sector has recently reported high number of incidents due to error, malware, and hacking\footnote{See for instance the, 2022 Data Breach Investigations Report, which shows that from the cases investigated, 67\% of them reported to have healthcare related issues - https://www.verizon.com/business/resources/reports/dbir/}. We therefore hypothesize that security risk can have a negative influence in trust in HVA.

\textbf{H10} Security Risk negatively influences Trust in HVAs.

\subsubsection{Privacy Risk (PR)} 
When users believe there is a potential risk for their data being misused or accessed by unauthorised parties, their trust may be affected~\cite{alraja2019effect,pitardi2021alexa,vimalkumar2021okay,foehr2020alexa}. In the healthcare domain, data can be successfully linked to individuals~\cite{hayden2013privacy,topol2019high} even if it has been anonymised and scrubbed of all identifiers~\cite{gymrek2013identifying}. As suggested in \cite{easwara2015privacy}, the more sensitive the information the more the likelihood of users having concerns when interacting with their voice assistants. Healthcare data is very sensitive~\cite{miner2016smartphone,abdi2021privacy}, so hypothesize that PR negatively affects trust in HVA.





\textbf{H11} Privacy Risk negatively influences Trust in HVAs.

\subsubsection{Perceived Substitution Risk (PSR)}
This factor is typically considered in healthcare technologies, and it captures the risk that traditional healthcare professionals may be replaced by technology in the future~\cite{fan2020investigating}, and that this could even lead to professional healthcare job loss~\cite{ongena2020patients,haan2019qualitative,nadarzynski2019acceptability}. There is also a risk that clinicians may lose skills that are increasingly carried out by technology (e.g. around diagnosis), resulting in increased errors when complex cases are escalated for human review~\cite{jutzi2020artificial}. All of this risk of substitution has been shown to affect users trust in AI technologies applied to healthcare in general~\cite{nelson2020patient}, so we hypothesize this to be also the case in HVA.

\textbf{H12} Perceived substitution risk negatively influences Trust in HVAs.

\subsection{Intention to Use}
As introduced in Section~\ref{sec: trust and VA}, the relationship between trust and adaption has been extensively studied. Across technologies, there seems to be a clear connection between users trusting and adopting technology~\cite{alhogail2018improving,lu2011dynamics,sharma2020consumer,liao2019understanding,pitardi2021alexa,al2022understanding,nasirian2017ai,lee2021role}. This has also been shown for the case of general-purpose voice assistants~\cite{nasirian2017ai,lee2021role,liao2019understanding,pitardi2021alexa,al2022understanding}. One of the proxies to attitudes towards adoption is \emph{intention to use} (or intention to adopt)~\cite{liao2019understanding,chao2019factors,al2022understanding}. In fact, previous research did show a strong connection between behavioural intentions and actual use of specific technologies~\cite{chao2019factors}. In accordance to this, we use in this paper intention to use as a proxy for adoption and hypothesize that: 



\textbf{H13} Trust in HVAs positively influences users' intention to use them.

\section{Methodology}
\label{methodology}
In order to test our hypotheses, we followed a quantitative methodology, by designing and administering a questionnaire with items measuring the factors introduced before as well as trust in HVAs and users' intention to use HVAs, and analyzing the results using Partial Least Squares Structural Equation Modelling (PLS-SEM). 
Our study was approved by our institution's IRB.



\subsection{Survey Development}

We created a questionnaire that contains three parts, which can be found in Appendix~\ref{appendix: questionnaire-part1}, \ref{appendix: item}, and \ref{appendix: post-survey questions}. The first part is a description of HVAs, so participants could understand what we mean by them. The second part includes the questions regarding the model and hypothesis introduced in the previous section in the form of items representing each of the factors, which were measured using a five-point Likert scale, with answer choices ranging from `strongly disagree' (1) to `strongly agree' (5). To ensure reliability and validity, the items were adapted from existing scales for each of the factors considered~\cite{mclean2019hey,gursoy2019consumers,fan2020investigating,cho2020will, hasan2020consumer,chi2021developing}. To do this, items were rephrased where appropriate to fit the study context (that is HVA). For example, one of the items for the factor \textit{Effort Expectancy} from \cite{chi2021developing} was rephrased from ``\emph{Using AI devices takes too much of my time}'' to ``\emph{Using healthcare voice assistants would take too much of my time}''. Every set of items for each factor was taken from the same source, so that all items measured the same underlying construct. 
The third and final part of the questionnaire includes three post-survey questions about users' perceptions of future use of HVAs, organisations they trust to design HVAs, and expectations for building trusted HVAs. 

To make sure that the questionnaire was effective and that participants understood the questions, we conducted two pilot studies with a total of 30 participants each. The data from the pilots was used only to improve the final survey and excluded from the final analysis. The pilots helped us refine the items, e.g. we were able to identify that some of the reversely-coded items (which we reversed for data quality purposes, as explained later) were confusing, and we adequately rephrased them. 






\subsection{Model Evaluation}
\label{scale-evaluation-method}

PLS-SEM was used for the statistical analysis of the second part of the questionnaire (items). This is because PLS-SEM has been found to be appropriate when the analysis is to test a theoretical framework from a prediction perspective and the structural model is complex and includes many factors~\cite{hair2019use}. 
In addition, PLS-SEM does not require large samples to perform well as some other SEM approaches do~\cite{hair2019use,hair2011pls}. 
We used the implementation of PLS-SEM in the SmartPLS 3.3 software~\cite{smartpls},
%
%
following the recommended two-step procedure~\cite{hair2011pls} of evaluating the measurement (outer) model first before estimating the structural (inner) model.

\subsection{Post-Survey Questions}

While statistical analysis of the model shows the factors that demonstrate statistical significance in influencing trust and intention to use, it is also important to understand participants' reasoning around the inclusion of certain specific factors and the exclusion of others. Consequently, we undertook a qualitative analysis to enhance our comprehension of the factors encompassed within the model as suggested by prior work~\cite{pitardi2021alexa,yin2013validity}. This qualitative exploration not only enabled us to delve deeper into the considered factors but also presented opportunities to identify additional factors that were not initially incorporated in our model, thus augmenting the existing body of literature in this domain.

For the third and final part of the questionnaire, responses to the open-ended questions were first converted into a text format, and then analyzed using thematic analysis following \citeauthor{braun2012thematic} \cite{braun2012thematic}, involving the 6 steps of familiarization, coding, generating themes, reviewing themes, defining and naming themes and writing up~\cite{braun2012thematic}. Individual coding was conducted by two researchers, and codes were recognized based on the ideas and feelings expressed by participants. Researchers then reviewed the codes created and discussed the results until reaching agreement. Similarly, the researchers then developed themes individually before coming together to discuss disagreements and converge on a final set of themes. 


\subsection{Data Collection}
\label{data-collection}
The final questionnaire was implemented using Qualtrics\footnote{\url{https://www.qualtrics.com}} and participants were recruited using Prolific\footnote{\url{https://prolific.co}}. All participants were voice assistant users aged 18 or older. The beginning of the questionnaire explained the study in detail, requested consent for data collection, and gathered information about participants' prior use of their voice assistant, including for healthcare purposes. 

To ensure a large enough sample, we followed two methods and retained the highest minimum number obtained. First, a minimum viable sample size was calculated using the “10-times rule”~\cite{thomson1995partial, hair2014primer}, which suggests the minimum sample should be 10 times the maximum number of arrowheads pointing at a latent variable (a formal term employed in PLS, synonymous with the 'factors' used in our study) anywhere in the PLS path model (12 arrows points to \emph{Trust in HVA}). This gave a minimum size of $10*12 = 120$ for each stage of data collection.

Second, to ensure that the sample also had sufficient statistical power to accurately detect true effects and minimize Type II errors we used the method proposed in \citeauthor{hair2009multivariate} \cite{hair2009multivariate}. This builds on Cohen’s power tables for least squares regression and relies on a table listing minimum required sample sizes based on three elements (the maximum number of arrows pointing at a latent variable, the significance level used, and the minimum $R^2$/smallest effect size that a researcher wishes to detect in the study in the model)~\cite{cohen1992power}. When the maximum number of arrows is 12, we need 181 observations to achieve a statistical power of 80\% for detecting $R^2$ values of at least 0.1 (with a 5\% probability of error). Therefore, 181 observations would be needed as a minimum for the study. 

\subsection{Data Quality and Reliability}
To ensure reliability and quality of the data collected, it is essential to employ rigorous measures for participant selection and questionnaire design. We employed \emph{three} widely recognized and frequently used measures in the questionnaire~\cite{hauser2016attentive,mason2012conducting,paas2018please,kim2019straightlining,peer2014reputation}: 

\begin{itemize}
    \item Recruitment of \emph{high-reputation} participants with at least 100 submissions and an approval rate of 95\% or more on the Prolific recruitment platform~\cite{peer2014reputation,such2017photo}.
    \item Use of two attention check questions to identify low-effort responses and filter those participants that failed either check from the analysis. These are essential for identifying inattentive respondents, promoting participants’ concentration and involvement, and improving the reliability of the data collected~\cite{hauser2016attentive,mason2012conducting, paas2018please}.
    \item Applying the Simple Non-differentiation Method after reverse-coding six questionnaire items~\cite{yan2008nondifferentiation, kim2019straightlining} to the responses in order to identify \emph{`straight lining'} by participants, excluding those who consistently selected the same response option across multiple questions (i.e., clicking in a straight line down the survey questions). 
\end{itemize}




	

\begin{table}[!h]
\centering
\caption{Demographics of the survey participants.}
\vspace{-10pt}
\small	
\begin{tabular}{lp{3.3cm}rr|lp{2.5cm}rr}
\toprule
               &                            & Par. & \% &                              &                                   & Par. & \% \\ \midrule
Gender   & Male    & 115  &38.3 & Employment & Full time  & 115  & 38.3 \\
 & Female & 182 & 60.7  &  & Part-time  & 36 & 12  \\
 & Prefer not to say & 3 &  1  & & Unemployed& 17 & 5.7   \\ \cmidrule{1-4}
Age   & 18-24   & 79   &  26.3 &  & Other & 14& 4.7\\
 & 25-34 & 116  & 38.7 &   & Retired/Homemaker & 16  &  5.3 \\
 & 35-44 & 61  &  20.3 &   & Prefer not to say & 102 &34    \\ \cmidrule{5-8} 
& 45-54  & 30  &   10  & Student& Yes   & 81  &27    \\
& 55-64  & 12  & 4 &    & No & 142     & 47.3   \\
 & 65 +     & 3  &  1  &     & Prefer not to say & 77   &    25.7        \\ \midrule
Assistant  & Amazon Echo/Alexa    & 142 &  47.3& Healthcare  & Total    & 147  &   49 \\
  used    & Amazon Echo/Alexa + others & 51  &  17 &      use of  & Information about illness/drug    & 71           &  23.7   \\
 & Google Home  & 75  & 25  &  assistant & Diagnosis based on symptoms  & 63 & 21\\
& Google Home + others  & 10  &3.3&   & Monitor \& manage chronic disease & 9  & 3 \\
 & Microsoft Invoke/Cortana   & 6   & 2 &                  & Treatment Plan   & 3   & 1 \\
 & Apple HomePod/Siri   & 8    &  2.7 &  & Others  & 1     &  0.3  \\
& Other & 8&  2.7 &                              &                                   &              &   \\
               \bottomrule
\end{tabular}
\label{tb:demographics}%
\end{table}

\section{Results}
\subsection{Participants}
Overall, 355 participants were recruited for the study. Of these, 43 were removed from the analysis for failing one of the two attention checks, and 12 participants were eliminated after straight lining. As a result, 300 participants were retained for data analysis, which is higher than the 181 participants we had established as minimum in Section~\ref{data-collection}. The breakdown of demographics for the participants is reported in Table~\ref{tb:demographics}, and the full dataset is available online at \url{https://osf.io/pdj3q/}.

\subsection{Model Evaluation}
\label{methods-scale-evaluation}

\subsubsection{Measurement Model Analysis}

\begin{table}[h!]
\centering
\caption{Convergent and discriminant validity results.}
\vspace{-10pt}
\scriptsize
\scalebox{0.85}{
\begin{tabular}{p{0.3cm}ccc|cccccccccccccc}
\toprule
& $\alpha$ & CR  & AVE  & A      & SI     & EE     & F      & IU & PCC  & PRSQ   & PU     & PR & SR & TA  & T & ST    & PSR \\
  \midrule
A    & 0.804            & 0.885                 & 0.719                            & 0.848  &        &        &        &                  &        &        &        &              &               &        &       &        &              \\
SI              & 0.853            & 0.900                   & 0.693                            & 0.437  & 0.833  &        &        &                  &        &        &        &              &               &        &       &        &              \\
EE               & 0.870             & 0.920                  & 0.794                            & -0.325 & -0.340  & 0.891  &        &                  &        &        &        &              &               &        &       &        &              \\
F                & 0.832            & 0.898                 & 0.746                            & 0.212  & 0.190   & -0.084 & 0.864  &                  &        &        &        &              &               &        &       &        &              \\
IU & 0.776            & 0.869                 & 0.689                            & 0.572  & 0.502  & -0.401 & 0.259  & 0.830             &        &        &        &              &               &        &       &        &              \\
PCC              & 0.761            & 0.862                 & 0.675                            & 0.574  & 0.486  & -0.427 & 0.152  & 0.619            & 0.822  &        &        &              &               &        &       &        &              \\
PRSQ             & 0.802            & 0.883                 & 0.717                            & 0.488  & 0.480   & -0.343 & 0.171  & 0.611            & 0.523  & 0.846  &        &              &               &        &       &        &              \\
PU               & 0.788            & 0.862                 & 0.611                            & 0.653  & 0.555  & -0.549 & 0.193  & 0.685            & 0.690   & 0.597  & 0.781  &              &               &        &       &        &              \\
PR     & 0.907            & 0.941                 & 0.842                            & -0.549 & -0.392 & 0.321  & -0.120  & -0.596           & -0.593 & -0.497 & -0.627 & 0.918        &               &        &       &        &              \\
SR   & 0.885            & 0.929                 & 0.813                            & -0.553 & -0.418 & 0.309  & -0.190  & -0.582           & -0.587 & -0.496 & -0.657 & 0.687        & 0.902         &        &       &        &              \\
TA               & 0.840             & 0.902                 & 0.754                            & 0.301  & 0.239  & -0.137 & 0.348  & 0.352            & 0.264  & 0.294  & 0.267  & -0.271       & -0.275        & 0.869  &       &        &              \\
T           & 0.836            & 0.891                 & 0.672                            & 0.682  & 0.546  & -0.497 & 0.236  & 0.736            & 0.749  & 0.651  & 0.736  & -0.731       & -0.730         & 0.342  & 0.820  &        &              \\
ST              & 0.795            & 0.880                  & 0.711                            & 0.556  & 0.402  & -0.337 & 0.222  & 0.573            & 0.554  & 0.486  & 0.636  & -0.515       & -0.518        & 0.242  & 0.675 & 0.843  &              \\
PSR     & 0.815            & 0.888                 & 0.726                            & 0.011  & 0.008  & -0.047 & -0.005 & 0.048            & 0.054  & 0.038  & 0.097  & -0.009       & -0.004        & -0.081 & 0.073 & -0.016 & 0.852          \\

\bottomrule
\end{tabular}}

\label{tab: ave CR}
\end{table}

We tested the internal consistency and discriminant validity of the proposed research model. The component reliability (CR) and average variance extracted (AVE) for each construct were examined to ensure that they met the threshold criteria for internal consistency.\footnote{CR is used to gauge the consistency of survey items in assessing a certain unobservable concept. It measures how well these items work together. In contrast, AVE determines how accurately these items represent the underlying concept, or essentially, how much they 'hit the target'.} The results are in Table~\ref{tab: ave CR}, which shows that the CR values are all above 0.70 and all AVE values are above 0.50. Therefore, there are no convergent validity issues~\cite{bagozzi1988evaluation}. This was also confirmed with traditional Cronbach's $\alpha$ measure,\footnote{Cronbach's $\alpha$ tells us how closely related a set of questions are as a group. Higher values (typically over 0.7) suggest the questions are effectively measuring the same concept.} where all $\alpha$ values in our scale are greater than 0.7 (between 0.713 and 0.923) as reported in the same table. Furthermore, to assess discriminant validity, Fornell and Larcker~\cite{fornell1981evaluating} suggested assessing the AVE values of each construct with the inter-correlation scores within the correlations table. They recommended that the AVE score must exceed the scores of the inter‐correlations. As it can be seen in Table~\ref{tab: ave CR}, the AVE values for all of the constructs are higher than the square of their inter-construct correlations,
thus confirming no discriminant validity issues~\cite{fornell1981evaluating}.

In terms of construct reliability, the outer loadings analysed in our study are summarised in Table~\ref{tab:factor_loading}, and all of the constructs have individual construct reliability values that are much larger than the preferred level of 0.7, which according to ~\citeauthor{hair2019use} indicates that the construct explains more than 50 per cent of the item’s variance, thus providing acceptable item reliability, except ST1 (0.479), SR2 (0.256), PR1 (0.410), A3 (0.569), PCC4 (0.357), PCC5 (0.365) and PSR2 (0.612), which were therefore removed and are not included in the table. 



\subsubsection{Structural Model Analysis} \label{sec:structure-model-analysis}

We then examined the model’s predictive capabilities and the relationships between the constructs. We first examined whether there were collinearity issues in the structural model. By calculating the variance inflation factor (VIF) values for the factors,\footnote{VIF measures how much the input factors in a model influence each other. In situations where factors are closely related, indicated by a high VIF.} all VIF values were below the minimum recommended of 3~\cite{kock2012lateral}, with values from 1.389 to 2.780, thus confirming that multi-collinearity was not violated.

\begin{table}[h!]
\centering
\caption{PLS-SEM Analysis Results: Hypotheses testing results.}
\vspace{-10pt}
\small
\scalebox{0.85}{
\begin{tabular}{l|llrrrc}
 \toprule
Factor Type&&Hypothesis & $\beta$ value & T Statistics &$p$ value & Supported?\\
\midrule

Functional Factor&H1  & Anthropomorphism -\textgreater Trust in HVA        & 0.076    & 2.030   & 0.042    & $\checkmark$  \\
&H2  & Effort Expectancy -\textgreater Trust in HVA        & -0.064    & 2.120   & 0.034   & $\checkmark$   \\
&H3  & Perceived Usefulness -\textgreater Trust in HVA    & 0.277   & 6.089   &$\textless$0.001   & $\checkmark$ \\
&H4  & Perceived Content Credibility -\textgreater Trust in HVA    & 0.159  & 4.782   &$\textless$0.001   & $\checkmark$ \\
&H5  & Perceived Relative Service Quality -\textgreater Trust in HVA   & 0.107  & 3.187   &0.001   & $\checkmark$ \\
Personal Factor&H6  & Familiarity -\textgreater Trust in HVA   & 0.028  & 1.154   &0.249   & x \\
&H7  & Technology Attachment -\textgreater Trust in HVA   & 0.038  & 1.341   &0.180   & x \\
&H8  & Social Influence -\textgreater Trust in HVA   & 0.023  & 0.751   &0.453   & x \\
&H9  & Stance in Technology -\textgreater Trust in HVA   & 0.111  & 3.568  &$\textless$0.001 & $\checkmark$\\
Risk Factor&H10  & Security Risk -\textgreater Trust in HVA   & -0.138  & 3.149  &0.002   & $\checkmark$   \\
&H11  & Privacy Risk -\textgreater Trust in HVA   & -0.173  & 5.135  &$\textless$0.001  & $\checkmark$   \\
&H12  & Perceived Substitution Risk -\textgreater Trust in HVA   & 0.032  & 1.116  &0.264  & x  \\
Trust Impact&H13  & Trust in HVA -\textgreater  Intention to Use  & 0.737  & 24.597  &$\textless$0.001  & $\checkmark$  \\
\bottomrule
\end{tabular}}

\label{tab: t-statistics}
\end{table}

Next, a bootstrapping with 5000 sub-samples in SmartPLS with a ``path'' weight scheme was performed. The results, as shown in Table~\ref{tab: t-statistics}, illustrate support for the majority of hypotheses and indicate the importance of the relationships between the constructs. Among the thirteen hypotheses we considered, all five functional and design factors (H1 - H5), two risk factors (H10, H11) and one personal factor (H9) showed  a significant influence on Trust in HVAs. In addition, H13 was proved, demonstrating that trust in HVA positively affects users' intention to use HVAs. Each of these supported hypotheses has a T-statistic greater than 2, indicating that that the path coefficient obtained is also statistically significant~\cite{hair2017updated,hair2019use}. 


\begin{figure}[h]
    \centering
    \includegraphics[width=\textwidth]{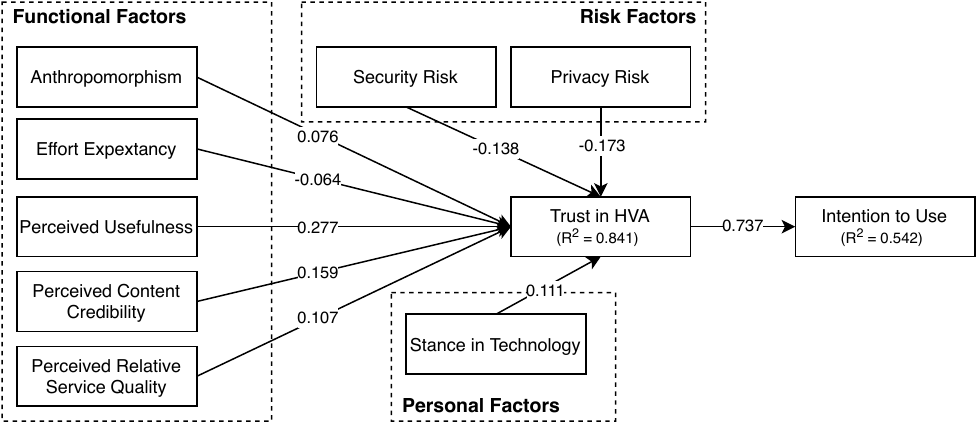}
    \vspace{-5pt}
    \caption{Final Research Model with Path Coefficients (Only significant paths are retained.)}
    \label{fig:numbered framework}
\end{figure}

Figure~\ref{fig:numbered framework} shows the final model only with significant paths. The $R^2$ was 0.841 and 0.542 for \textit{Trust in HVA, and Intention to Use}, respectively, which represents the explanatory power of the construct. We can say that, the \emph{Intention to Use} is \textbf{moderately} explained by \emph{Trust in HVA}, which in turn is \textbf{substantially} explained by its antecedent factors\footnote{As suggested by Hair et al.~\cite{hair2017updated,hair2019use}, $R^2$ values of 0.75, 0.50 and 0.25 can be considered substantial, moderate and weak.}. Finally, the bootstrapped Standardized Root Mean Square Residual (SRMR) is the most frequently used measure to determine goodness-of-fit while using SmartPLS~\cite{fernandes2021understanding}. In this study, the SRMR value is 0.053, below the 0.08 threshold, suggesting an adequate model fit.


\subsubsection{Controlling for HVA Use and Demographics} \label{sec:control}We also wanted to control for additional variables introduced by our sample of participants beyond the factors we hypothesized (as common in PLS-SEM analysis)~\cite{sohaib2021social}. Following guidance on the reporting of the use of control variables~\cite{atinc2012control}, we were particularly interested in our post-hoc analysis in the previous use of voice assistants for healthcare as well as other demographics reported by our participants (see Table~\ref{tb:demographics}). Although all our participants had experience using a voice assistant, not all of them were users of HVA (30.7\% used HVA). This was important for properly modelling intention to use HVAs, something not possible if all participants were already using them. However, it was also important to understand whether that may have influenced trust, so we added to the model a  control variable of previous HVA use on Trust in HVA. We also added control variables to capture the voice assistant they normally use (e.g. Amazon Alexa, Google Assistant, etc.) following~\cite{seymour2021exploring}, as well as general demographics like age~\cite{bailey2019systematic} and gender~\cite{doi:10.1080/15332861.2021.1927437} which may also have an influence in trust in technology in general. The resulting model (which is reported in the Appendix~\ref{appendix: post-hoc}) showed no significant differences in terms of the factors previously analyzed; those significantly and substantially influencing trust in HVA remained so, and those that were not significant remained so as well. In terms of the control variables, only gender was statistically significant, suggesting women to be more inclined to trust in HVAs (Sample Mean = 0.107, $p \textless$0.001).

\subsection{Post-survey Results}
From the qualitative analysis of open-ended questions, we elicited themes that helped us better understand and supplement our structural model: 1) \emph{Security and Privacy (associated with H10 and H11), 2) Anthropomorphism (associated with H2), 3) Other functional-related (associated with H2-5), 4) Financial Causes, 5) Discrimination Issues, 6) Develop Organisation, and 7) Substitution Risk (associated with H12)}. While the model had already identified the first three themes,  themes 4--6 emerged as \emph{novel} findings from the thematic analysis. Moreover, the factor, \emph{Perceived Substitution Risk}, which has not demonstrated statistical significance and therefore has been excluded from the model, is also not backed up by the results derived from the thematic analysis. The themes and codes can be found in Appendix \ref{appendix:themes and codes}. 62 responses were excluded due to the provision of non-substantive or irrelevant responses to the posed questions, e.g., "I don't know" or "I am not sure". 


\subsubsection{Security and Privacy (retained)}
Security and privacy was a common theme mentioned by participants when asked about the use of healthcare voice assistants, they have shared several concerns including data misuse. In particular P113 (\emph{Amazon Echo/Alexa\&Google Home, none}) mentioned, "I would feel uncomfortable discussing personal issues that Amazon employees could hear". When asked about the future development of healthcare AI assistants, several mentioned improving security features;  P7 (\emph{Amazon Echo/Alexa\&Google Home, Seek diagnosis results \& Ask illness or drug information}) said "security improvement" and P178 (\emph{Amazon Echo/Alexa, Ask illness or drug information}) said "A good approach regarding the security of the data collected". Although some participants are open to the idea of adopting HVAs, there are still some concerns regarding trust.

\subsubsection{Anthropomorphism (retained)}
Participant responses helped us understand the reasons why factors, such as Anthropomorphism, were ultimately retained in the model. More specifically, 24.4\% participants felt that a trustworthy HVA should firstly meet the basic ability to \emph{"understand the user's description and have a natural conversation (P27, Google Home, Seek diagnosis results \& Ask illness or drug information)}" with them, and secondly, at a higher level, \emph{"they should be real and communicate like a human, less robot. This requires them to have more human-like features or customised voices. (P60, Amazon Echo/Alexa, Ask illness or drug information)}". In addition to this, considering that discussing a health condition is a serious and potentially emotionally draining matter, participants resonated strongly with the idea that a trustworthy HVA should be able to show compassion and empathy, with four 
participants mentioning that \emph{"Although they are honest, it should have the empathy that doctors have when they say things like someone is going to die. (P16, Amazon\&Microsoft Cortana, Seek diagnosis results\&monitor a chronic disease; P56, Amazon Echo/Alexa, Seek diagnosis results; P103, Amazon Echo/Alexa, Seek diagnosis results \& Ask illness or drug information; P223, Amazon Echo/Alexa\&Apple Homepod, none)}".

\subsubsection{Other functional-related (retained)}
Other functionality related factors were also mentioned such as the accuracy of the diagnosis provided by the HVA, P176 \emph{(Google Home, Seek diagnosis results \& Ask illness or drug information)} mentioned "Quality and accuracy of physical diagnosis, particularly in medical areas where people will feel uncomfortable seeing a doctor about". Others have mentioned the knowledge of HVAs and their ability to provide accurate responses, P252 \emph{(Google Home, none)} "I would want to make sure they are consistent and knowledgeable". Usability was also been mentioned, P235 \emph{(Amazon Echo/Alexa, Ask illness or drug information)} "easy ways to access health services".

\subsubsection{Financial Causes (new)} Price seems to be what will sway users to use and rely on HVAs particularly in places where healthcare is not provided as a public, free service. 
A participant points out that \emph{"As for the cost, they should consider a broader group of users to make it affordable for everyone.(P242, Google Home, Seek diagnosis results \& Ask illness or drug information\&monitor a chronic disease)"}. For this, the majority of those impacted seems are not wealthy individuals, like one participant said: \emph{"HVAs that give healthcare access to under-privileged populations would win more people support."} Apart from that, users are willing to trust and use HVAs as long as they are not set up to make a profit (\emph{"An AI that is not influenced by profit motives but can supply cost information when requested (P209, Amazon Echo/Alexa, Ask illness or drug information)}".).

\subsubsection{Discrimination Issues (new)} Another factor to increase trust and adoption would be the \textbf{reducing discrimination} in the health care delivery process. Some users (2\%) mentioned that they had received discrimination in traditional healthcare settings for different reasons and had experienced discomfort during their visits to  doctors, so they wished that the HVAs could \emph{"focus only on the facts (P235, Amazon Echo/Alexa, none)}". They were very confident in the usage of HVAs and one participant expressed that \emph{"HVAs should help reduce the bias that many doctors have towards patients so that a real diagnosis could happen and help the patient (P292, Google Home, Seek diagnosis results \& Ask illness or drug information)}".

\subsubsection{Develop Organisation (new)} 
\begin{figure}[h]
    \centering
\includegraphics[scale = 0.45] {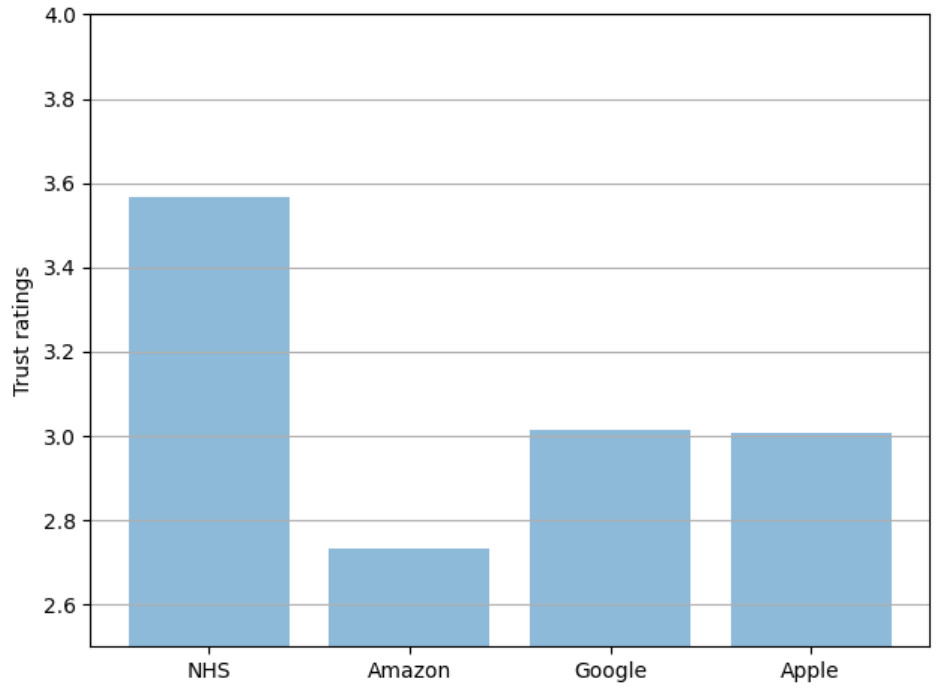}
\vspace{-10pt}
    \caption{Most trusted organizations to develop HVA.} 
    \label{fig:trust_organisation}
    \vspace{-5pt}
\end{figure}

Finally, an interesting result comes from the analysis of the \textbf{trusted organisations} that develop HVAs. In asking participants who would develop a HVA, it was concluded that the UK's NHS (i.e. the UK's National Health Service) had a much higher reputation among participants than big companies such as Google and Apple, while Amazon supporters were the least of these options (see Figure~\ref{fig:trust_organisation}). Beyond the above parties, the most common type participants recommended were private tech companies, followed by hospitals and academic institutions.

\subsubsection{Substitution Risk (excluded from the final model)}. With regards to reasons why some factors were not retained in the model, the most common theme was about substitution risk. In particular, 22.3\% participants do not feel that doctors will be replaced by HVAs in the future. On the contrary, they believe that HVAs should complement the work of doctors, \emph{"I think it should be a tool used by doctors to help them, and not something used to replace them (P43, Google Home, none})." The interesting thing is when HVAs work alone, especially on cases of serious illness, participants don't trust them to work well. But they trust the HVAs if their information and decisions can be monitored by a doctor (\emph{"It is more reliable that they work alongside doctors, and decisions they made should be further validate by a real doctor (P300, Amazon Echo/Alexa, Ask illness or drug information)}.").

%

\section{Discussion}
\label{discussion}


We now summarize the contributions of our model and then compare it to previous research on trust in general-purpose voice assistants. After this, we discuss the main implications for HVAs and provide recommendations for developers and researchers who want to explore HVAs.




This study makes critical theoretical and practical contributions by firstly explaining user trust in HVAs through factors captured in three important aspects \emph{(personal, functional, and risk)}. These dimensions are recognized as pivotal elements in the overarching framework of human-computer interaction, particularly in the context of fostering collaborative relationships between humans and AI entities~\cite{jirotka2005collaboration,10.1145/3500868.3559450,park2019identifying}. Specifically, the model provided a statistically based assessment of the degree to which the various factors influence trust, in addition to testing which factors increase or decrease it. According to the results, in each of the categories of personal, functional and risk factors, at least one factor was shown to have an impact on trust. This is a crucial step towards a holistic model to design trustworthy HVAs with a concrete understanding of the antecedents that foster trust. 
%
From the factors we considered, the most significant ones that affect user trust in HVAs are usefulness, information credibility, service quality, and security/privacy, exploring how they might be designed for in current and future products. We learned potential reasons why certain factors were retained in the model and were meaningful in explaining trust, while others were not (for instance because participants thought substitution was not really a risk in how they viewed HVA should work). In addition, we learned more about the users and tried to discover aspects that have not been taken into account in terms of trust (for instance who builds the HVA, if it helps mitigate discrimination and if it is affordable). 

Our research, especially its results, offers profound insights that resonate with the broader society and, more specifically, the intricate healthcare sector. This field, characterized by its complexity, epitomizes the essence of multi-stakeholder collaboration, a theme central to the ethos of the CSCW community~\cite{fitzpatrick2013review,lai2021human,park2019identifying,park2017beyond}. By delving into the dynamics of trust and adoption, we provided AI product providers with a clearer roadmap for product refinement, emphasizing areas that foster enhanced collaboration~\cite{jirotka2005collaboration,10.1145/3500868.3559450}. This collaborative lens can also help policymakers and third parties understand patient-centric concerns, notably privacy and data safety. Furthermore, when compared with general-purpose voice assistants, our findings accentuate the distinct collaborative and social complexity of the healthcare domain. In doing so, our work not only addresses societal needs but also aligns with and contributes to the core values and interests of the CSCW community, emphasizing the significance of cooperation.

\subsection{Comparing Trust Between HVA and General-purpose Voice Assistants}

Following the views expressed in previous literature~\cite{montague2009empirically,miner2016smartphone,abdi2021privacy}, this study reveals the necessity of studying trust in the healthcare context, especially compared with trust in general-purpose technologies, in this case HVA in comparison with general-purpose voice assistants. 
To better compare the results of this study with previous work on trust in \emph{general-purpose} voice assistants, we first summarize the outcomes of the factors being validated in literature on trust in general-purpose voice assistants in Table~\ref{tab:compare-with-general}. 
Based on these previous works and our results, we highlight four main differences in terms of trust in HVA and trust in general-purpose voice assistants. 

First, we considered factors that have not been studied in prior work on general-purpose voice assistants but that, as explained throughout this paper, have been shown to be key in the context of healthcare technologies (perceived content credibility, perceived relative service quality and perceived substitution risk). In this regard, we show that two functional factors, perceived content credibility and perceived service quality do in fact influence trust in HVAs, but that perceived substitution risk does not. According to the qualitative results that, 
the current use of HVAs is regarded as a complement to existing healthcare services rather than a replacement, something we discuss further in the next section on practical insights.

\begin{table}[!h]
    \centering
    \small
     \caption{Support for factors influencing trust in general-purpose voice assistants. N/C means not considered in the literature. 
     }
     \vspace{-10pt}
    \begin{tabular}{c|lll}
    \toprule
     Category &  Factor & Supported & Rejected  \\
     \midrule
       Functional  & Anthropomorphism & \cite{chi2021developing,pitardi2021alexa,foehr2020alexa} & -\\
       &Effort Expectancy& \cite{alhogail2018improving,chi2021developing,pitardi2021alexa} & -\\
        &Perceived Usefulness& - & \cite{pitardi2021alexa}\\
        &Perceived Content Credibility& N/C & N/C\\
        &Perceived Relative Service Quality& N/C & N/C\\
        \midrule
        Personal & Familiarity& \cite{alraja2019effect,chi2021developing} & -\\  & Technology Attachment& \cite{chi2021developing} & -\\
       & Social Influence& \cite{alhogail2018improving,chi2021developing} & -\\ 
        & Stance in Tech& \cite{chi2021developing} & -\\
        \midrule
 Risk & Security Risk & \cite{alhogail2018improving,alraja2019effect} & -\\
  & Privacy Risk & \cite{alraja2019effect,vimalkumar2021okay} & \cite{pitardi2021alexa}\\
  &Perceived Substitution Risk & N/C & N/C\\
      \bottomrule  
    \end{tabular}
    \label{tab:compare-with-general}
\end{table}

Second, previous work on general-purpose voice assistants has not found a direct link between perceived usefulness and trust~\cite{pitardi2021alexa}. In contrast, this study shows that for HVAs, perceived usefulness is actually the factor that contributes the most towards substantially explaining trust. The reason for this may be that HVAs have a specific purpose, a hypothesis which is consistent with previous work on healthcare technology that also found perceived usefulness to influence trust~\cite{peng2020patient}. On the other hand, general purpose assistants tend to have more varied purposes that are not found in the healthcare context, such as hedonic benefits and entertainment~\cite{pitardi2021alexa}.


Third, among the personal factors, only Stance in Tech proved to be directly linked to trust in HVA. Even the social influence factor, where a device owner's relationship may affect user perceptions in general settings~\cite{10.1145/3359161}, seems not to hold true in the distinct environment of healthcare. This may be because what leads people to trust VAs in general, i.e. familiarity and attachment to the platform~\cite{alraja2019effect,chi2021developing}, being recognised or influenced by others~\cite{alhogail2018improving}, are not as important in the perceptions of users as other factors, such as the usefulness and quality evidenced in the results of our model. It may also be that not all participants actually have people in their social networks that use HVAs or they may not be familiar with it. However, when controlling in the model for HVA use (recall that half of our participants used HVA) it was shown as not significant. 

Finally, prior work on general-purpose voice assistants is inconclusive as to the effect of privacy risks on trust, with some studies suggesting that privacy risks influence trust~\cite{alraja2019effect,vimalkumar2021okay} and others suggesting that it does not~\cite{pitardi2021alexa}. We show that perceived privacy risks are very important for trust in HVAs, being the second most significant factor in the final model. One explanation for this is that healthcare data is known to be very sensitive in general~\cite{miner2016smartphone,abdi2021privacy}, and previous work on privacy and voice assistants in particular found that users considered the flows of this data across the voice assistant ecosystem as the least appropriate~\cite{abdi2021privacy}. 

\subsection{Implications Unique to Healthcare Voice Assistants}


\paragraph{Enhancing usefulness in HVAs through participatory design.} Perceived usefulness plays the most important role in the user's trust in HVAs. This suggest that engagement with patients throughout the process of HVA development is crucial, from ideation through design, implementation, and evaluation. A common and well established means of achieving highly usable solutions in healthcare and other areas of HCI is participatory design methods~\cite{spinuzzi2005methodology,harrington2020design}, and beyond this it is important to engage with the communities that are the target of an intervention throughout the design and development process. It has been shown, for instance, that as a response to the lack of usability of some diabetes technologies for many patients, users have begun hacking their way into commercial devices to better serve their needs (e.g. do-it-yourself automated insulin delivery systems~\cite{braune2022open}). Beyond usefulness as an objective it is also important that users \emph{perceive} HVAs to be useful, particularly as potential users may have had negative experiences with voice assistants in the past.

\paragraph{Balancing service quality and roles in HVAs.} The quality of service relative to what a health professional would offer also contributes substantially and significantly to trust in HVA. When considering this, there are two key roles HVA could play that may make it easier or harder to offer a quality of service that is higher relative to clinical staff. The first of these roles would be HVAs as a \textit{replacement} for in-person care, implying a challenging goal of matching the diagnostic, analytical, and social/communication skills. It has been recommended that HVAs should be more compassionate~\cite{tanioka2021development} and exhibit more non-verbal communication skills and emotional reactions~\cite{aeschlimann2020communicative}. This also requires that HVAs are able to personalise responses to different people; behaving like a human healthcare professional requires empathy when facing serious cases, as well as observing the user's personality traits in other consultation situations. This is an area where development is required. At present, research reports mixed results and a lack of testing in real-world clinical environments~\cite{gille2020we,vollmer2018machine,nagendran2020artificial}. The second, and potentially easier, role for HVAs would be as a \textit{complement} to existing healthcare services. In these scenarios, HVAs might be restricted to triage, giving limited advice on less complex conditions and handing off other cases to human practitioners. This is a much simpler task, lowering the cost of service delivery whilst being much easier to adequately perform relative to a medical professional (as an example, matching the advice given by a clinician for a common cold is much easier than for type 2 diabetes). 
In the qualitative results, it becomes evident that the ideal scenario for patients involves the integration of healthcare AI but supervised and regulated by physicians. This not only exemplifies the need for harmonious collaboration between doctors and cutting-edge technology for the betterment of patient care, as highlighted in~\cite{lai2021human, fitzpatrick2013review, park2019identifying}, but also underscores the importance of human-AI interaction in fostering trust between patients and their physicians.


\paragraph{Privacy complexities in HVAs and design choices.} While privacy is a key concern in general-purpose voice assistants, the introduction of healthcare data presents additional complexities. Many research prototypes and currently available HVAs are built with or on top of general-purpose commercial voice assistants~\cite{sezgin2020scoping}. The security and privacy of voice assistants in general is an active area of research---see~\cite{edu2020smart,seymour2023systematic} for a systematic review on the topic--- and several issues have been revealed, including the extensive collection of sensitive user data, unauthorized access and questionable practices by third-parties and skills~\cite{edu2021skillvet,guo2020skillexplorer,seymour2023voice,bispham2022leakage,edu2022measuring}, along with inadequacies in the development and application of privacy policies~\cite{liao2020measuring,edu2022measuring}. Researchers are actively working on a range of solutions, such as mechanisms for providing user with meaningful verbal consent~\cite{seymour2023legal,chalhoub2024useful} and automated systems to identify policy breaches~\cite{young2022skilldetective,edu2021skillvet}. Additionally, significant efforts are being directed towards creating models that aid users in making privacy decisions consistent with their preferences~\cite{zhan2022model, zhan2023privacy} and crafting explanations that incorporate elements of trust and privacy, which are intended to diminish user concerns and strengthen trust in VAs~\cite{seymour2023ignorance,such2017privacy}.
Moreover, how users' perceive security and privacy of these underlying platforms is very important for HVAs, as users consider healthcare-related data more sensitive and prefer to share less information with voice assistants about health~\cite{abdi2021privacy}. As the information they share with them is often protected by additional regulation (such as HIPAA in the U.S.), the security of those HVAs is therefore highly dependent on the security of the voice assistants on top of which they are built. To accommodate this, certain platforms such as Alexa have additional requirements for health-related skills\footnote{\url{https://developer.amazon.com/en-US/docs/alexa/custom-skills/requirements-for-hipaa-eligible-skills.html}} --- skills are the name given to the third-party functionality that can be added to Alexa to add capacities to it. Furthermore, attitudes towards health information privacy have notably shifted over time, particularly in the context of chronic illnesses such as diabetes~\cite{o2013non}. For instance, the emergence of health-related complications within a patient's medical condition may result in a diminished emphasis on privacy. Additionally, the assumption of new responsibilities, such as the care of children afflicted with chronic illnesses, may prompt individuals to reassess their privacy considerations. Given the evolving concerns, especially in chronic illnesses like diabetes, it's imperative to address these in the design of voice assistants handling sensitive health data.

An alternative option is to create standalone HVAs (instead of on top of general-purpose voice assistants), mitigating the risks and negative perceptions related to voice assistant platforms. However, the vast amount of data and expertise required in order to create the speech recognition and NLP processing services required to operate a successful voice assistant means that, in practice, these services will likely be provided by third parties, who may even be the same organisations providing general-purpose voice assistants. While there may be differences between, e.g., creating an HVA skill to run on top of Amazon Alexa and creating a standalone HVA using Amazon Lex\footnote{A ``fully managed artificial intelligence (AI) service with advanced natural language models to design, build, test, and deploy conversational interfaces in applications'' \url{https://aws.amazon.com/lex}}, that difference may be difficult to communicate to users.


\paragraph{Trust dynamics and the importance of compliance in development.} An often overlooked point in previous research, is who are expected to make the HVAs? For some mainstream technology companies (Amazon, Google, and Apple), user trust in VAs when they are used for no specific purpose correlates with trust in their parent company, and Amazon has a higher level of "trust reputation" than Google ~\cite{seymour2021exploring} in the minds of users. However in terms of providing health-related services, Amazon was less trusted than Google/Apple in our results. This may be explained by the fact that platforms collaborate with different partners that provide/develop health applications, and the strength of the platform in regulating the compliance of these applications with existing privacy and security policies can also affect user trust building. Evidence suggests that breaches are common in Amazon Alexa skills for health care, with 86.36\% missing disclaimers when providing medical information, and 30.23\% storing users' physical or mental health data without approval~\cite{shezan2020verhealth}. 
Beyond big tech companies, public health organizations such as the UK's NHS are trusted more to develop HVA. 
This is probably because the NHS system is largely funded by the UK government and is under the jurisdiction and oversight of the Department of Health\footnote{\url{https://www.nhs.uk/}}.
Meanwhile, given that the services provided by the NHS are free to UK citizens, this reinforces users' belief that the purpose of NHS is to help people, improve the healthcare environment and save lives, rather than earning money. This is in line with the argument made by users that they trust non-profit organisations more as for-profit companies often do not have the most stringent track record in protecting data and may sell these data without user awareness. Therefore, handing the company the keys to people’s private health information raises red flags. Furthermore, each country will have its own clinical guidelines, and as we mentioned before, some countries often have additional regulations governing the use of health information (HIPAA in the U.S.). In this regard, each healthcare system would potentially suggest different organisations who would develop/manage such an assistant. This might be redundant for VAs that provide common functions (e.g. ordering a taxi, managing the To-do List), but it is necessary for HVAs. Moreover, deeper integration between large companies with established program frameworks and hospitals with specialist physicians and public credibility should be encouraged as well.

\paragraph{Discrimination and cost.} Last but not least, other factors identified in our post-survey questionnaire should also be considered for trust in HVA in practice, such as the potential for HVA to be more or less discriminatory than health professionals and the impact this may have on users. 
Participants seem to think that HVA would discriminate and have less biases and preconceptions than health professionals. However, 
it is well-established that AI-based systems may not necessarily be any more \emph{neutral}, and they can actually be as discriminatory as, or even more than, human decision-makers \cite{ferrer2021bias,nuenen2022intersectional}. Beyond benefiting from the general efforts towards fairer AI \cite{binns2018fairness}, this will also require HVAs developers to have a deeper understanding of discrimination in the medical industry and thoroughly evaluate future HVAs in the context of these ethical issues. With regards to financial issues, the cost of HVAs may also need to be considered.

\subsection{Limitations and Future Work}
\label{limitation_future}

This study has some limitations. First, our model is a significant step towards a holistic model to explain factors that influence trust in HVA. However, there are still many factors in all levels of designing an HVA that can impact on trust and should be taken into account (e.g., factors that were newly identified in qualitative results, financial causes, bias, and develop organisation). Future work should focus on this and continue to extend our model. It is crucial to delve deeper into the influential factors that affect this field of study. Considerations such as the systematic categorisation of functional factors, the improved capacity for rectifying communication errors and failures as evidenced in \cite{cho2020role}, and their subsequent effect on trust, demand attention. Additionally, privacy and security concerns associated with general voice assistants~\cite{lau2018alexa,malkin2022can,meng2021owning,mcmillan2019designing,seymour2023ignorance,10.1145/3359161}, their application within the healthcare context, and their potential influence on trust also necessitate investigation. These represent significant opportunities for future work within this research domain. Second, 
trust may evolve as time goes by and users have more chances to interact with HVA. 
Therefore, 
longitudinal studies should also be considered in the future to consider factors that contribute to trust building over time, such as the type of relationship that could be formed between a user and an HVA, with recent studies having explored the relationships that users develop with general-purpose voice assistants~\cite{seymour2021exploring}. 
Concurrently, as discussed in section 6.2, developers should continually update their efforts to ensure functional ease and user safety mechanisms. Given that attitudes, including privacy concerns and needs, can evolve over time~\cite{o2013non}, it's imperative to develop and offer services that provide enduring benefits to users.
Third, we focused on users of HVAs as \emph{patients}, but another interesting demographic to consider would be healthcare professionals, who 
may also utilize HVAs to facilitate their work, e.g. by allowing doctors to have voice interactions with electronic health records~\cite{kumah2018electronic} or supporting the tasks of home health aides~\cite{bartle2022second}. 
This would complement our study from the `two sides' of potential HVA use. 

\section{Conclusion}
\label{conclusion}

Our work is the first one to examine different functional, personal and risk factors that affect trust in healthcare voice assistants and user intention to use these technologies.  We find that anthropomorphism, perceived usefulness, effort expectancy, content credibility, and relative service quality, together with security risk, privacy risk and stance in technology substantially explain trust in HVA. In turn, trust in HVA moderately explains intention to use HVA. Based on these results, we derive some practical insights and recommendations to help design and develop next-generation HVAs that can promote user trust and encourage adoption.

\begin{acks}

We thank CSCW’s anonymous reviewers, Ruba Abu-Salma, and Hana Kopecka for their constructive comments on previous drafts of this paper. 
This research was partially funded by EPSRC under grant \emph{SAIS: Secure AI assistantS} (EP/T026723/1) and the INCIBE's strategic SPRINT (Seguridad y Privacidad en Sistemas con Inteligencia Artificial) project with funds from the EU-NextGenerationEU through the Spanish government's Plan de Recuperación, Transformación y Resiliencia.

\end{acks}

\bibliographystyle{ACM-Reference-Format}
\bibliography{sample-base}


\appendix


\section{Questionnaire - Part One: Scenario Description} \label{appendix: questionnaire-part1}
After the participant has chosen to consent to starting the questionnaire, a detailed scenario description is provided to help them understand the subject of the study (what we mean by HVAs):

\begin{center}
 
\emph{`Imagine that you have an AI-based voice assistant at home which take care of your health, and you can seek help without going out to see a GP. You can issue commands directly to the voice assistant or have conservation with third-party skills that deployed on the voice assistants. It can take care of both your physical and mental health like a GP, such as: making a swift and effective diagnosis based on the symptoms that appear; assessing the likelihood of a certain illness; discussing and developing a treatment plan with you; helping you monitor and manage your chronic disease. Explain to you the test results such as blood test, x-ray diagnosis, etc. Combining with your own experience with voice assistants, please answer the following questions, we want your honest thoughts.`} 
\end{center}

\section{Questionnaire - Part Two: Item List}\label{appendix: item}

The second part of the survey is to let participants rate each of the items listed below using a 5-point likert scale that goes from from ‘strongly disagree’ (1) to ‘strongly agree’ (5).


\noindent \textbf{Functional Factors}\\
\noindent \textit{Anthropomorphism (A) - adapted from~\cite{pitardi2021alexa}:}
   
    \noindent A1. When I interacting with healthcare AI voice assistants, it would feel like someone is present in the room. \\
    A2. I feel that interactions with healthcare AI voice assistants would be similar to those with a human.\\
    A3. When communicating with healthcare AI voice assistants, I would feel like I am dealing with a real person.\\
    A4. I think I would communicate with healthcare AI voice assistants in a similar way to how I would communicate with other people.

\noindent \textit{Effort Expectancy (EE) - adapted from~\cite{chang2017user,chi2021developing}: }

    \noindent EE1. I feel using healthcare AI voice assistants would take too much of my time. \\
    EE2. I feel working with healthcare AI voice assistants would be difficult and I would not understand how to use them.\\
    EE3. I feel it will take me too long to learn how to interact with healthcare AI voice assistants.

\noindent \textit{Perceived Usefulness (PU) - adapted from~\cite{cho2020will}:}

    \noindent PU1. I feel healthcare AI voice assistants would help me be more efficient when dealing with health-related issues.\\
    PU2. I feel healthcare AI voice assistants would be useful when dealing with health-related issues. \\
    PU3. I feel healthcare AI voice assistants could meets my needs.\\
    PU4. I feel healthcare AI voice assistants would be able to do everything I would expect in order to take care of my health.

\noindent\textit{Perceived Content Credibility (PCC) - adapted from~\cite{cho2020will}:}

\noindent Please indicate how well the following adjectives represent the healthcare AI voice assistant, from 1 = describes very poorly to 5 = describes very well:

    \noindent PCC1. Accurate \\
    PCC2. Complete\\
    PCC3. Expert\\
    PCC4. Consistent \\ 
    PCC5. Authentic 

\noindent\textit{Perceived Relative Service Quality (PRSQ) - adapted from~\cite{fan2020investigating}:}

    \noindent PRSQ1. I think the accuracy of information provided by a healthcare AI voice assistant would be higher than that of the average doctors.\\
    PRSQ2. I think question answering and diagnosis speed of healthcare AI voice assistants would be faster than that of the average doctors.\\
    PRSQ3. I think Healthcare AI voice assistants would provide a clearer and more understandable service delivery process.

\noindent\textbf{Personal Factors}

\noindent \textit{Stance in Technology (ST) - adapted from~\cite{chi2021developing}:}

\noindent ST1. I usually keep an eye on emerging products using AI technology, especially those that will be beneficial to my health.\\ 
ST2. I always try out emerging products using AI technology earlier compared to others especially if they will be beneficial to my health. \\ 
ST3. In general, I am willing to accept new emerging products using AI technology, especially if they will be beneficial to my health. \\
ST4. If I heard about an emerging product using AI technology, especially that is beneficial to my health, I would look for ways to use it.

\noindent \textit{Familiarity (F) - adapted from~\cite{chi2021developing}:}

\noindent F1. I know a lot about healthcare AI voice assistants. \\
F2. I have much knowledge about healthcare AI voice assistants.  \\
F3. I am more familiar than the average person regarding healthcare AI voice assistants.

\noindent \textit{Technology Attachment (TA) - adapted from~\cite{chi2021developing}:}

\noindent TA1. I feel that AI technology is a part of my identity .\\ 
TA2. I identify strongly with the use of AI technology. \\ 
TA3. Using AI digital technology says a lot about who I am.

\noindent \textit{Social Influence (SI) - adapted from~\cite{alhogail2018improving}:}

\noindent SI1. Using healthcare AI voice assistants would be a status symbol in my social networks (friends, family, co-workers).\\
SI2. People in my social networks who would utilize healthcare AI voice assistants have more prestige than those who wouldn't. \\
SI3. People whose opinions I value would encourage me to utilize healthcare AI voice assistants. \\
SI4. People in my social networks who would utilize artificial intelligence such as healthcare AI voice assistants have a high profile.

\noindent\textbf{Risk Factors}

\noindent \textit{Security Risk (SR) - adapted from~\cite{cho2020will}:}

\noindent SR1. I think it would be risky to interact with a Healthcare AI voice assistant.\\
SR2. I would be concerned if I had to deal with my health-related issues via a healthcare AI voice assistant. \\
SR3. There would be much uncertainty associated with my interactions with a healthcare AI voice assistant.

\noindent \textit{Privacy Risk (PR): - adapted from~\cite{cho2020will}:}

\noindent PR1. I am concerned that a healthcare AI voice assistant would collect too much information about me.\\
PR2. I am concerned about who might access my personal information I had given to a healthcare AI voice assistant.\\
PR3. I feel that healthcare AI voice assistants would misuse my personal information.\\
PR4. There would be a large potential loss associated with providing personal information to healthcare AI voice assistant.

\noindent \textit{Perceived Substitution Risk (PSR) - adapted from~\cite{fan2020investigating}:}

\noindent PSR1. I think that healthcare AI voice assistants are likely to replace doctors in the future.\\
PSR2. I think using healthcare AI voice assistants for a long time would make doctors dependent on them.\\
PSR3. I think the rise and development of healthcare AI voice assistants would likely lead to the unemployment of some doctors. \\
PSR4. I think using healthcare AI voice assistants for a long time would decrease doctors’ own ability.

\noindent\textbf{Trust and Intention to Use}

\noindent \textit{Trust in HVAs (T) - adapted from~\cite{cho2020will}:}

\noindent T1. I feel Healthcare AI voice assistants would be interested in my well-being.\\
T2. I feel like healthcare AI voice assistants would be truthful.\\
T3. I would characterize healthcare AI voice assistants as honest. \\
T4. I feel like a healthcare AI voice assistant would be sincere.

\noindent \textit{Intention to Use (IU) - adapted from~\cite{pitardi2021alexa}:}

\noindent IT1. It is likely that I will use my voice assistant for Healthcare in the future.\\
IT2. I intend to use my voice assistant for healthcare frequently.\\
IT3. I expect to continue using my voice assistant for healthcare in the future. 


\section{Questionnaire - Part Three: Post-survey questions} \label{appendix: post-survey questions}
The last part of the survey contains both closed questions and open-ended questions that aims to understand more about how users perceive healthcare voice assistants. The responses to these questions were converted into a text format that could be analysed using thematic analysis methods.




\noindent Q1: Would you consider to use a healthcare voice assistant in the future? 


- if No, then why you may NOT wish to use a healthcare AI assistant in the future?

\noindent Q2: I would trust the following organisations to make a healthcare voice assistant? not at all(1) - very much(5)

- The NHS (on another platform)

- Amazon

- Google

- Apple

- Please briefly describe any other organizations you trust to develop healthcare AI assistants (if any)

\noindent Q3: What aspects should be enhanced in terms of developing a trustworthy voice assistant in healthcare?

\section{Thematic analysis - themes and codes}\label{appendix:themes and codes}

Table~\ref{tab: qualitative-codes} displays the themes and the corresponding codes investigated from participants’ response to the post-survey questions.

\begin{table}[!h]
\centering
\footnotesize
\caption{Themes and codes of thematic analysis.}
\label{tab: qualitative-codes}
\vspace{-10pt}
\begin{tabular}{lll}
\toprule
Relation to the model & Themes                    & Codes                           \\
\toprule
Retained                & Security and Privacy      & Reduce general privacy concern  \\
                        &                           & Data and usage transparency     \\
                        &                           & Ensure data storage security    \\
                        &                           & Obey regulations such as HIPAA  \\
                        &                           & Prevent data selling            \\
                        \hline
Retained                & Anthropomorphism          & Provide natural conversation    \\
                        &                           & Empathy                         \\
                        &                           & Compassionate                   \\
                        &                           & Customize voice and care      \\
                        &                           & More human-like features        \\
                        \hline
Retained                & Other functional-related  & Higher accuracy                 \\
                        &                           & Higher usability                \\
                        &                           & Knowledgeable                   \\
                        &                           & Timely                          \\
                        &                           & Ease of use                     \\
                        &                           & Help support avaliable          \\
                        &                           & Reliable information \& advice  \\
                        \hline
Excluded                & Substitution Risk & Assist instead of replace       \\
                        &                           & Supervised by doctors           \\
                        &                           & Help contact doctors            \\
                        \hline
Newly found             & Financial Causes          & Affordable to more people       \\
                        &                           & Special care to underprivileged \\
  &                           & Not designed for profit \\                        
                        \hline
Newly found             & Discrimination Issues      & Eliminate discrimination        \\
                        &                           & Treat patients equally          \\
                        \hline
Newly found   & Develop Organisation    & NHS>Google$\approx$Apple>Amazon         \\
 &      &  Private tech companies       \\
&      &  Hospital \& academic institutions       \\
                        \bottomrule
\end{tabular}
\end{table}

\section{Item Loadings} \label{appendix:item loading}
The final version of item loadings are shown in the Table~\ref{tab:factor_loading}. As can be seen from the table, the loadings are much larger than the preferred level of 0.7, which according to ~\citeauthor{hair2019use} indicates that the construct explains more than 50 per cent of the item’s variance, thus providing acceptable item reliability, except ST1 (0.479), SR2 (0.256), PR1 (0.410), A3 (0.569), PCC4 (0.357), PCC5 (0.365) and PSR2 (0.612), which were therefore removed and are not included in the table.

\begin{table}[h!]
    \centering
    \small
    \renewcommand{\arraystretch}{1.0}
        \caption{Item Loading for Each Factor}
        \vspace{-10pt}

    \begin{tabular}{p{3.4cm}lc|p{3.4cm}lc}
    \toprule
      Factor  &  Item &  Loading & Factor  &  Item &  Loading \\
      \midrule
Perceived   Substitution Risk (PSR) & PSR1 & 0.906 & Anthropomorphism (A)                      & A1    & 0.861 \\
                                    & PSR3 & 0.830  &                                           & A2    & 0.872 \\
                                    & PSR4 & 0.817 &                                           & A4    & 0.810  \\
Perceived Usefulness (PU)           & PU1  & 0.760  & Social Influence (SI)                   & SI1   & 0.864 \\
                                    & PU2  & 0.760  &                                           & SI2   & 0.713 \\
                                    & PU3  & 0.843 &                                           & SI3   & 0.857 \\
                                    & PU4  & 0.760  &                                           & SI4   & 0.885 \\
Trust in HVA (T)                    & T1   & 0.787 & Effort Expectancy (EE)                    & EE1   & 0.879 \\
                                    & T2   & 0.803 &                                           & EE2   & 0.899 \\
                                    & T3   & 0.869 &                                           & EE3   & 0.894 \\
                                    & T4   & 0.818 & Perceived Relative Service Quality (PRSQ) & PRSQ1 & 0.871 \\
Intention to Use (IU)               & IU1  & 0.833 &                                           & PRSQ2 & 0.845 \\
                                    & IU2  & 0.824 &                                           & PRSQ3 & 0.822 \\
                                    & IU3  & 0.833 & Security Risk (SR)                        & SR1   & 0.926 \\
Technology Attachment (TA)          & TA1  & 0.913 &                                           & SR2   & 0.863 \\
                                    & TA2  & 0.909 &                                           & SR3   & 0.914 \\
                                    & TA3  & 0.777 & Privacy Risk (PR)                         & PR2   & 0.923 \\
Stance in Technology (ST)    & ST2 & 0.766 &                                           & PR3   & 0.923 \\
                                    & ST3 & 0.889 &                                           & PR4   & 0.907 \\
                                    & ST4 & 0.869 & Perceived Content Credibility (PCC)       & PCC1  & 0.838 \\
Familiarity (F)                     & F1   & 0.828 &                                           & PCC2  & 0.853 \\
                                    & F2   & 0.893 &                                           & PCC3  & 0.772 \\
                                    & F3   & 0.869 &                                           &       &   \\
    \bottomrule
    \end{tabular}
  
    \label{tab:factor_loading}
\end{table}

\section{Post-hoc Analysis of the Model} \label{appendix: post-hoc}
As mentioned in Section~\ref{sec:control}, out of interest in whether the use of HVA and user-reported demographics also influenced trust in HVAs, we ran a second model
which added the control variables to capture users' general demographics like \emph{age} and \emph{gender}, as well as whether they have \emph{experience} in using voice assistant for healthcare purpose, and the voice assistant (\emph{device type}) they normally use (e.g. Amazon Alexa, Google Assistant, etc.). Note that, from the demographics of the survey participants (see Table~\ref{tb:demographics}), 3 of the 300 participants reported that they were not willing to state their gender status, and in order to include the gender variable as a binary variable in the model, we evaluated the models after removing the data associated with these three participants. To summarise, the first model is the model \emph{without} these three participants and \emph{without} the control variables, and the second is the model \emph{without} these three participants but \emph{with} the control variables. The numbers in brackets in the table \ref{table1} are the results of the second model. Finally, the resulting model showed no significant differences in terms of the factors previously analyzed; those significantly and substantially influencing trust in HVA remained so, and those that were not significant remained so as well. In terms of the control variables, only \textbf{\emph{gender}} was statistically significant, suggesting women to be more inclined to trust in HVAs (Sample Mean = 0.107, $p \textless$0.001).

\begin{table}[!h]
\footnotesize
\centering
\caption{PLS-SEM Analysis Results: Hypotheses testing results (O: Original Sample Mean, M:Sample Mean, STDEV: Sandard Deviation). The numbers in brackets are results for models with control variables added.}  
\vspace{-10pt}
\scalebox{0.85}{
\begin{tabular}{llcccccc}
\toprule
    &                                                               Hypotheses & O & M & STDEV & T statistics& $p$ Value & Supported? \\ \hline
H1  & Anthropomorphism -\textgreater{}Trust in HVA                   & 0.076(0.073)        & 0.076(0.073)    & 0.038(0.036)               & 2.006(2.015)             & 0.045(0.044) &  $\surd$($\surd$)     \\
H2  & Effort Expectancy -\textgreater{}Trust in HVA                  & -0.059(-0.065)      & -0.059(-0.065)     & 0.030(0.030)                      & 1.977(2.147)             & 0.048(0.032) &    $\surd$($\surd$)        \\
H3  & Perceived Usefulness -\textgreater{}Trust in HVA               & 0.278(0.281)        & 0.278(0.282)    & 0.046(0.047)               & 6.011(5.956)             & 0.000(0.000) &     $\surd$($\surd$)       \\
H4  & Perceived Content Credibility -\textgreater{}Trust in HVA      & 0.159(0.156)        & 0.159(0.156)    & 0.034(0.035)               & 4.669(4.504)             & 0.000(0.000) &   $\surd$($\surd$)         \\
H5  & Perceived Relative Service Quality -\textgreater{}Trust in HVA & 0.106(0.104)        & 0.105(0.103)    & 0.034(0.033)               & 3.125(3.108)             & 0.002(0.002) &   $\surd$($\surd$)         \\
H6  & Familiarity -\textgreater{}Trust in HVA                        & 0.029(0.030)        & 0.029(0.031)    & 0.024(0.024)               & 1.2(1.215)               & 0.230(0.225) &    x(x)        \\
H7  & Technology Attachment -\textgreater{}Trust in HVA              & 0.039(0.024)        & 0.041(0.025)    & 0.029(0.028)               & 1.376(0.864)             & 0.169(0.388) &    x(x)        \\
H8  & Social Influence -\textgreater{}Trust in HVA                   & 0.024(0.016)        & 0.025(0.015)    & 0.032(0.032)               & 0.761(0.505)             & 0.447(0.614) &   x(x)         \\
H9  & Stance in Technology -\textgreater{}Trust in HVA               & 0.112(0.100)        & 0.11(0.099)     & 0.032(0.032)               & 3.549(3.082)             & 0.000(0.002) &     $\surd$($\surd$)       \\
H10 & Security Risk -\textgreater{}Trust in HVA                      & -0.138(-0.107)      & -0.138(-0.108)  & 0.044(0.045)               & 3.104(2.399)             & 0.002(0.016) &    $\surd$($\surd$)        \\
H11 & Privacy Risk -\textgreater{}Trust in HVA                       & -0.175(-0.158)      & -0.173(-0.156)  & 0.033(0.033)               & 5.224(4.840)             & 0.000(0.000) &   $\surd$($\surd$)         \\
H12 & Perceived Substitution Risk -\textgreater{}Trust in HVA        & 0.032(0.015)        & 0.03(0.015)     & 0.029(0.028)               & 1.087(0.516)             & 0.277(0.606) &    x(x)        \\
H13 & Trust in HVA -\textgreater{}Intention to Use                   & 0.738(0.738)        & 0.738(0.739)    & 0.031(0.030)               & 24.155(24.293)           & 0.000(0.000) &    $\surd$($\surd$) \\
\hline
 & \textbf{Gender -\textgreater Trust in HVA }                  & (0.108)        &(0.107)    & (0.027)               & (4.005)           & \textbf{(0.000)} &  \textbf{($\surd$)}  \\
 & Age -\textgreater Trust in HVA                 & (0.009)        &(0.009)    & (0.023)               & (0.375)           & (0.708) &   (x)  \\
 & Experience -\textgreater Trust in HVA                 & (0.002)        &(0.003)    & (0.026)               & (0.093)           & (0.926) &   (x)  \\
 & Alexa -\textgreater Trust in HVA                 & (0.002)        &(0.003)    & (0.026)               & (0.093)           & (0.926) &  (x)   \\
 & Google -\textgreater Trust in HVA                 & (0.026)        &(0.052)    & (7.595)               & (0.003)           & (0.997) &   (x)  \\
 & Others -\textgreater Trust in HVA                 & (0.023)        &(0.106)    & (7.596)               & (0.003)           & (0.998) &   (x)  \\
\bottomrule
\end{tabular}}

	\label{table1}  
\end{table}

\received{January 2023}
\received[revised]{July 2023}
\received[accepted]{November 2023.}

\end{document}